\documentclass[showpacs,pre,amssymb,amsmath]{revtex4}
\usepackage{graphics}
\usepackage{psfrag}
\begin{document}
\title{Analysis of path integrals at low temperature: \\
 Box formula, occupation time and ergodic approximation}
\author{S{\'e}bastien Paulin, Angel Alastuey,
Thierry Dauxois\thanks{Thierry.Dauxois@ens-lyon.fr}}
\affiliation{Laboratoire de Physique, UMR-CNRS 5672, ENS Lyon, 46
All\'{e}e d'italie, 69364 Lyon c\'{e}dex 07, France}
\date{\today}
\begin{abstract}
We study the low temperature behaviour of path integrals for a simple
one-dimensional model. Starting from the Feynman-Kac formula, we
derive a new functional representation of the density matrix
at finite temperature, in terms of the occupation times of
Brownian motions constrained to stay within boxes with finite sizes.
From that representation, we infer a kind of ergodic approximation,
which only involves double ordinary integrals.
As shown by its applications to different confining potentials,
the ergodic approximation turns out to be quite efficient, especially
in the low-temperature regime where other usual approximations fail.
\end{abstract}
\maketitle
\section{ Introduction}

\noindent The knowledge of the density matrix at finite temperature
$T$ for few interacting particles, is of paramount importance for
studying equilibrium properties of quantum many-body systems.
Indeed, such density matrices occur in low-fugacity expansions,
while they are also key ingredients in quantum Monte Carlo
simulations. In this context, the Feynman-Kac (FK) representation of
the density matrix in terms of path
integrals~\cite{FeyHib,SimSchRoe} is particularly useful. On the one
hand, it has been used for deriving exact (analytical) expressions
for simple models~\cite{Kle}. On the other hand, beyond the
well-known Wigner-Kirkwood expansion~\cite{Wig,Kir,LanLif} around
the classical limit, various approximations, non-perturbative in
$\hbar$, have been introduced within that framework: for instance,
the celebrated semi-classical approximation~\cite{FeyHib} (widely
used in numerous situations), the variational approach of Feynman
and Kleinert~\cite{FeyKle}, or the renormalized Wigner-Kirkwood
scheme.

\bigskip

\noindent The FK representation is described in Sec.~\ref{quantumformalism}.
For the sake of simplicity, we consider
the simple case of a single particle in one dimension submitted to a
stationary confining potential.
The mean spatial extension of the brownian paths, which intervene in the
FK representation, is controlled by the thermal de Broglie wavelength of
the quantum particle. At high temperatures, since that extension vanishes,
exact asymptotic expansions, as well as direct numerical
estimations, become rather straightforward. At the analytical level,
Wigner-Kirkwood $\hbar^2$-expansions around the
classical limit, which have been derived long ago in
other frameworks, are easily recovered. On the computational side,
functional integration over paths can be replaced, within a
good accuracy, by ordinary integrals over a discrete set of parameters,
\textsl{via} the use of either free propagators in Trotter formula~\cite{Kra}
or simple modelizations of brownian paths~\cite{Pre}.
However, at low temperatures, similar calculations loose accuracy,
while they also become rather time-consuming. Moreover, the corresponding
asymptotic structure of the density matrix, provided by the
groundstate contribution in its spectral expression, does
not clearly emerge from its FK representation. In fact, in
that temperature regime, the average extension of paths diverge,
whereas the main contribution to the functional integral
arises from paths with a finite
extension, of the order of the localization
length of the ground state: contributions of such minority paths are
significantly different from the average contribution.
Most of our knowledge is negative, i.e. tells us which trajectories are
not important~\cite{bogojevic}, so, in some sense, we
will try to have a positive attitude~\cite{raffarinade}.

\bigskip

\noindent In this paper, we derive a new functional representation of
the density matrix, which is more suitable than the genuine FK formula
for tackling the low-temperature regime (see Sec.~\ref{rewritingFeynamanKac}).
The starting central observations are described in Sec. III.A. First, in
the FK functional integral, only marginal paths with
finite extensions (i.e. large deviations with respect to the average)
contribute when $T$ goes to zero. Second, many of such paths
with quite different jagged shapes, provide similar contributions, mainly
determined by the corresponding local occupation times. Thus, it
is quite natural to collect paths into sets defined by their
spatial extension and their local occupation times. That procedure
allows us to transform exactly FK representation~(\ref{FK}) into the
so-called box formula~(\ref{2B3}). That formula is defined
\textsl{via} the introduction of paths constrained to stay in a box
with size $\ell$ and characterized by an intermediate flight time $s$
(in $\beta \hbar$ units with $\beta=1/k_BT$). It involves
a double ordinary integral over $\ell$ and $s$, combined to functional
integrals over local occupation times associated with
the constrained paths. As required, box formula~(\ref{2B3})
provides a better understanding of the low-temperature behaviour of
the density matrix than FK representation~(\ref{FK}). When $T$ vanishes,
leading contributions obviously arise from typical sizes $\ell$
much smaller than de Broglie wavelength. In a forthcoming paper,
we will argue how groundstate
quantities emerge from box formula~(\ref{2B3}),
by using scaling properties of the probability distribution function (PDF)
of occupation times  at low temperatures.

\bigskip

\noindent Beyond its conceptual interest for understanding
low-temperature behaviours of path integrals,
box formula~(\ref{2B3}) also allows us to derive new approximations.
This is illustrated in Section~\ref{constrainedaverages}, where we present
the so-called ergodic approximation. That approximation results from
the truncation to first order of cumulant expansions of the
functional averages. It amounts to replace each local occupation
time associated with a given constrained path, by its average over the
PDF of such occupation times. This can be viewed as some kind
of ergodic hypothesis, because the (imaginary) time average
of the potential experienced by
the particle along that path, is then replaced by a spatial average with
a measure defined by the previous mean occupation-time.
Ergodic expression~(\ref{est1_1b}) involves only an ordinary double integral
over $\ell$ and $s$. The key ingredient, namely the mean occupation-time
(in units of $1/\ell$), depends on three dimensionless parameters. Using
its low and high temperature behaviours (derived analytically), we
propose simple tractable expressions for that quantity, which turn out
to be quite accurate at any temperature.

\bigskip

\noindent Section~\ref{applications}
is devoted to the applications of the ergodic approximation to
various simple forms of the confining potential. First,
we determine the asymptotic analytical forms of the approximate
density matrices at both low and high temperatures. When $T$ diverges,
Wigner-Kirkwood expansion is partially recovered. When $T$ vanishes,
the main features of the exact behaviours are well reproduced.
The approximate density matrices then do factorize as a product of
a Boltzmann factor associated with a given energy, times a
function of position only: this provides satisfatory
approximate expressions for the groundstate energy and wavefunction.
Second, numerical calculations are performed at finite intermediate
temperatures. As expected from the previous analytical results,
the ergodic approximation turns out to be quite reasonably accurate
over the whole range of considered temperatures, and
discrepancies with (numerically) exact results (in part inferred from the
spectral representation) do not exceed a few percent. Moreover,
it significantly improves over the well-known semi-classical
approximation, which completely fails at low temperatures
(except of course for the harmonic potential where semi-classical
calculations become exact).

\bigskip

\noindent Further applications and extensions of our approach 
will be presented elsewhere. In particular,
the ergodic approximation can be applied to unconfining potentials, and it
can be straightforwardly extended to higher dimensions and to systems with
several particles. Furthermore, other approximations might be
easily introduced by starting from box formula as argued at the end 
of Section III.

\section{Path integral framework}\label{quantumformalism}

\noindent In this Section, we first define the model and then introduce
the FK representation of the corresponding density matrix.
Next, we  briefly recall the efficiency of FK formula for
describing the high-temperature regime, and we argue about
its drawbacks for analyzing low-temperature behaviours.

\subsection{The model}

\noindent We consider a quantum non-relativistic particle of mass $m$ in
one dimension $z$, submitted to a confining potential $V$, i.e
$V(z) \rightarrow \infty$ when $|z| \rightarrow \infty$. Its
Hamiltonian reads
\begin{equation}
H=-\frac{\hbar^2}{2m} \frac{\mathrm{d}^2}{\mathrm{d}
z^2} + V(z).\label{hamiltonian}
\end{equation}
Introducing inverse temperature $\beta=1/(k_B T)$ with Boltzmann constant
$k_B$, we define the Gibbs operator
\begin{equation}\label{scrho_dst}
\rho= \exp{\left[-\beta H \right]} ,
\end{equation}
the matrix elements of which define the so-called density matrix
$\rho(x,y,\beta)=\langle x \left| \rho \right| y \rangle$.
The partition function
\begin{equation}
Z(\beta)=\text{Trace}\,\left[ \, \rho\, \right]
=\int_{-\infty}^{+\infty} \mathrm{d} z\, \rho(z,z,\beta)
\end{equation}
is well behaved thanks to the confining nature of potential $V$
(in other words, integral over $z$ does converge because $\rho(z,z,\beta)$
decays sufficiently fast at large distances). The (normalized)
probability density to find the particle at position $x$ then reduces to
\begin{equation}\label{expressionrho}
\Psi \left(x,\beta \right) =
\frac{\rho(x,x,\beta)}{Z(\beta)}.
\end{equation}

\bigskip

\noindent For further purposes, it is also convenient to
introduce the spectral representations of density matrix and partition
function, which read
\begin{equation}\label{scrho_dst_element}
\rho(x,y,\beta) = \sum_{k=0}^{+\infty} \phi_k(x) \phi_k^*(y)
\exp{\left(-\beta E_k \right)}
\end{equation}
and
\begin{equation}
Z(\beta) = \sum_{k=0}^{+\infty} \exp{\left(-\beta E_k \right)}.
\label{partspec}
\end{equation}
In such representations, $E_k$ is the $k$-th eigenvalue of
$H$, while $\phi_k$ is its associated eigenfunction which satisfies
Schr\"odinger equation
\begin{equation}
[-\frac{\hbar^2}{2m} \frac{\mathrm{d}^2}{\mathrm{d}
z^2} + V(z)]\phi_k(z)=E_k\phi_k(z).
\label{schro}
\end{equation}

\subsection{Feynman-Kac formula}

\noindent Path integrals were first introduced for representing the
matrix elements of the evolution operator $\exp(-iHt/\hbar)$ associated with 
Schr{\"o}dinger equation \cite{FeyHib}. It was soon realized that
a similar path integral representation for matrix elements of
Gibbs operator at inverse temperature $\beta$ can be inferred
\textsl{via} the formal substitution $t \rightarrow -i\beta \hbar$.
The latter representation for $\rho(x,y,\beta)$ involves an integral over
all paths $\omega(\cdot)$ going from $\omega(0)=x$ to $\omega(\beta \hbar)=y$
in a time $\beta \hbar$. Moreover, the corresponding integrand is
the exponential factor $\exp[-S(\omega(\cdot))/\hbar]$, where
$S(\omega(\cdot))$ is the action associated with path $\omega(\cdot)$.
That path integral is proved to be mathematically well defined for a
large class of potentials. Its convergence is indeed
ensured, roughly speaking, by the Gaussian-like decay of the
kinetic part of $\exp[-S(\omega(\cdot))/\hbar]$ for large paths
(on the contrary, path integrals for $\exp(-iHt/\hbar)$ are, in general,
ill-defined because they involve sums of oscillating phase factors).

\bigskip

\noindent The genuine path integral representation of $\rho(x,y,\beta)$
can be rewritten within the parametrization
$\omega(u) = (1-s)x + sy + \lambda_D \xi(s)$, where $s=u/(\beta \hbar)$ is
the dimensionless time in $\beta \hbar$ units, while $\xi(s)$ is
a Brownian bridge satisfying boundary conditions $\xi(0)=\xi(1)=0$
and $\lambda_D=(\beta \hbar^2/m)^{1/2}$ is the de Broglie wavelength.
On the one hand, the kinetic part of $\exp[-S(\omega(\cdot))/\hbar]$ then provides
normalized Wiener measure $\mathcal{D}_W(\xi)$ associated with the
Brownian bridge process. That Gaussian measure is entirely defined by its
first two moments
\begin{eqnarray}
&&\int_\Omega  \mathcal{D}_W(\xi)\ \xi(s)=0,\label{loopsproperty2}\\
&&\int_\Omega  \mathcal{D}_W(\xi) \ \xi(s_1)\,\xi(s_2)=\min(s_1,s_2)
\left(1-\max(s_1,s_2) \right).\label{loopsproperty3}
\end{eqnarray}
On the other hand, the potential part of $\exp[-S(\omega(\cdot))/\hbar]$
reduces to the Boltzmann factor associated with time
average of $V$ along path $(1-s)x + sy + \lambda_D \xi(s)$. The resulting
so-called Feynman-Kac formula reads~\cite{SimSchRoe}
\begin{equation} \label{FK}
\rho(x,y,\beta)=
\frac{\exp [-(x-y)^2/(2\lambda_D^2)]}{\sqrt{2\pi}\lambda _D}
\int_\Omega \mathcal{D}_W (\xi)
\exp \left(- \beta \int_0^{1} \mathrm{d} s\, V \left(
(1-s)x + sy + \lambda_D \xi(s) \right) \right)
\end{equation}
where  $\Omega=\left\{ \xi(\cdot) \right\}$ is the
infinite set of realizations of the Brownian bridge process.

\bigskip

\begin{figure}[ht]
\begin{center}
\resizebox{0.8\textwidth}{!}{\includegraphics{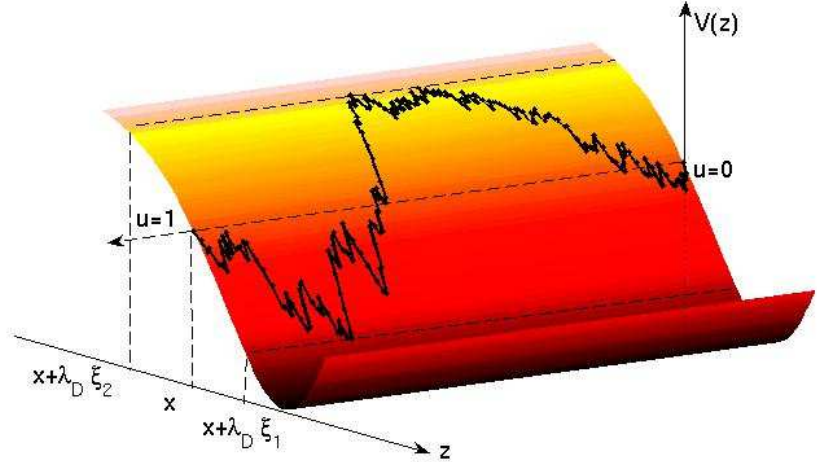}}
\caption{Typical path which starts and ends at position
$x$, along which particle experiences potential~$V$. Dashed lines
indicate the two edges of a fictitious box
determined by the extremal deviations $\xi_1$ and $\xi_2$ of the
path. Typical sizes of $\xi_1$ and $\xi_2$ are controlled by $\lambda_D$}
\label{deflambdadebroglie}
\end{center}
\end{figure}

\noindent In the Feynman-Kac formula~(\ref{FK}), the Wiener measure is
intrinsic to Brownian motion and does not depend on any physical
parameter. Expression~(\ref{loopsproperty3}) of the covariance implies that the
typical extension of a  Brownian bridge is
always of order $1$. Mass $m$ of the particle, as well as Planck's constant
$\hbar$ only intervene in the de Broglie wavelength $\lambda_D$. That length
controls the size of quantum fluctuations, as shown by
specifying (\ref{FK}) to diagonal elements, i.e.
\begin{equation} \label{FKd}
\rho(x,x,\beta)=
\frac{1}{\sqrt{2\pi}\lambda _D}\int_\Omega \mathcal{D}_W (\xi)
\exp \left(- \beta \int_0^{1} \mathrm{d} s\, V \left(
x + \lambda_D \xi(s) \right) \right).
\end{equation}
In time-average $\int_0^{1} \mathrm{d} s\, V \left(
x + \lambda_D \xi(s) \right)$,
particle experiences the potential around position $x$ on a
length scale obviously determined by $\lambda_D$, as illustrated in
Figure~\ref{deflambdadebroglie}.

\bigskip

\noindent Besides its interest for analytical or numerical calculations
(see below), Feynman-Kac representation is also quite useful for
mapping a quantum system into a classical one~\cite{Gin,Cor,Mar}. Indeed,
formula~(\ref{FKd}) can be reinterpreted in terms of a classical
extended object, often called loop or polymer, defined by
its position $x$ and its internal shape  parametrized by
$\lambda_D \xi(\cdot)$. Then, the r.h.s. of (\ref{FKd}) is
nothing but the statistical average of a
Boltzmann factor for the loop over its all possible shapes distributed
according to Wiener measure. That interpretation, which holds for
any quantum system with an arbitrary number of particles, is
the key starting point of above mapping.

\subsection{High and finite temperatures}

\noindent At high temperatures, the de Broglie wavelength is small,
so only paths which remain close to reference position $x$ contribute significantly to $\rho(x,x,\beta)$ in (\ref{FKd}). Therefore,
the time-average of $V$
along path $x+\lambda_D \xi(s)$ can be performed by replacing
$V \left( x+\lambda_D \xi(s)\right)$ by its Taylor series around
$V \left(x\right)$
\begin{equation}
V \left( x+\lambda_D \xi(s)\right) = V(x) + \sum_{n=1}^{\infty}
{(\lambda_D \xi(s))^n \over n!} {\mathrm{d}^n V \over \mathrm{d} x^n}
\label{TaylorV}
\end{equation}
(we assume that $V$ is infinitely differentiable).
If we keep only the first term in (\ref{TaylorV}), which does not depend
on $\xi(s)$, we immediately recover the familiar classical form of
$\rho(x,x,\beta)$
\begin{equation} \label{classical}
\frac{1}{\sqrt{2\pi}\lambda _D}
\exp[- \beta V(x)],
\end{equation}
where we have used normalization condition
$\int_\Omega \mathcal{D}_W (\xi)=1$. Since $\lambda_D$ is proportional
to $\hbar$, quantum corrections to (\ref{classical})
are easily generated by a (formal) cumulant expansion of
contributions associated with terms $n \geq 1$ in (\ref{TaylorV}).
The resulting moments of Wiener measure are then readily computed by
Wick's theorem which applies thanks to Gaussian character of
$\mathcal{D}_W (\xi)$. This allows us to retrieve the well-known
Wigner-Kirkwood expansion in powers of $\hbar^2$
\begin{eqnarray}\label{WK}
\rho(x,x,\beta) = \frac{e^{-\beta V(x)}}{\sqrt{2 \pi}\lambda_D}
\left[ 1 -{\beta^2 \hbar^2 \over 12m}\frac{\mathrm{d^2} V}{\mathrm{d}x^2}
+ {\beta^3 \hbar^2 \over 24m} \left( \frac{\mathrm{d} V}{\mathrm{d}x}\right)^2
+ ...\right].
\end{eqnarray}
Expansion~(\ref{WK}) involves only even powers of $\hbar$, because
all odd moments of gaussian Wiener measure vanish by virtue of
(\ref{loopsproperty2}). That expansion is expected to be only asymptotic.
Present derivation within FK representation turns out to be much easier than
usual approaches within Wigner-distribution formalism (see e.g
ref.~\cite{LanLif}). It might be used for calculating higher-order
terms beyond $\hbar^2$-correction (terms up to order
$\hbar^6$ have been already determined in the literature~\cite{WK6}).
Moreover, it clearly emphasizes that (\ref{WK}) is appropriate when
the de Broglie wavelength is much smaller than the typical length of
variation of the potential. Then, truncation of (\ref{WK}) up
to a few terms should provide accurate values of the density matrix, as
checked below.

\bigskip

\noindent When the temperature decreases, the de Broglie wavelength increases,
so Wigner-Kirkwood expansion is no longer applicable. At the analytical
level, exact evaluations of the corresponding path integrals are not
accessible in general, except for some simple models~\cite{Kle}.
Non-perturbative effects in $\hbar$ can be partially accounted for
through various methods, among which the celebrated
semi-classical approximation described further. The reliability of such
methods remains questionable since they do not involve
any small control-parameter like (\ref{WK}). Nonetheless, for
numerical purposes, their accuracy may be satisfactory, at
least at not too low temperatures.

\bigskip

\noindent If analytic informations on the density matrix
become rather difficult to extract from FK representation, accurate
values can be still readily obtained at finite temperatures
\textsl{via} direct numerical estimations of FK path integrals. A first
possible route starts with the (exact) convolution relation
\begin{equation}
\rho(x,x,\beta) = \int \prod_{i=1}^N \mathrm{d} z_i
\prod_{i=0}^N \rho(z_i,z_{i+1},\beta/(N+1))
\label{convolution}
\end{equation}
with $z_0=z_{N+1}=x$. For a sufficiently large value of $N$, each
matrix element $\rho(z_i,z_{i+1},\beta/(N+1))$ can be evaluated within
FK formula~(\ref{FK}) by neglecting the corresponding Brownian
deviation $\lambda_D \xi(s)/(N+1)^{1/2}$ in the argument of~$V$. The resulting
time average $\int_0^{1} \mathrm{d} s
V \left((1-s)x + sy \right)$ can be itself
replaced, with a similar accuracy, by $V((x + y)/2)$ (middle-point
convention) for instance. The resulting approximate expression
\begin{equation}
\int \prod_{i=1}^N \mathrm{d} z_i
\prod_{i=0}^N \frac{\exp [-(N+1)(z_i-z_{i+1})^2/(2\lambda_D^2)]}
{\sqrt{2\pi}\lambda _D/(N+1)^{1/2}}
\exp [-(\beta/(N+1) \sum_{i=0}^N V((z_i + z_{i+1})/2) ]
\label{convapprox}
\end{equation}
turns out to be quite accurate for rather small values of $N$, provided that
$\lambda_D/(N+1)^{1/2}$ is smaller than the typical length of
variation of the potential. We stress that (\ref{convapprox}) becomes exact
when $N \rightarrow \infty$. In fact, continuous path integration with
Wiener measure may be precisely defined in the framework of that limit
process.

\bigskip

\noindent A second route for direct numerical computations relies on the
introduction of a set of $N$ simple functions $e_n(s)$ ($n=1...N$) with $e_n(0)=e_n(1)=0$, such that they define a basis for Brownian bridges
$\xi(s)$ in the limit $N \rightarrow \infty$ \cite{Pre}. Numerical
evaluations of (\ref{FKd}) are then carried out by replacing the
functional integration over paths by ordinary integrals over
coefficients $a_n$ arising in the decomposition $\xi(s)=a_ne_n(s)$.
That method provides satisfactory results by using only a few
number $N$ of basis functions \cite{Pre}. Similarly to the previous
convolution method, it becomes exact when $N \rightarrow \infty$.

\subsection{Low temperatures}

\noindent When temperature goes to zero, $\lambda_D$
diverges, so most paths explore a rather large region not restricted
to the neighbourhood of reference position $x$. A direct analysis of FK
expression~(\ref{FKd}) then becomes quite cumbersome. In particular,
$\rho(x,x,\beta)$ must behave as
\begin{equation} \label{low_temp_assympt}
\rho(x,x,\beta) \stackrel{\beta
\to +\infty}{\sim} | \phi_0(x) |^2 \exp{\left[-\beta E_0\right]},
\end{equation}
which immediately follows from spectral representation of density
matrix~(\ref{scrho_dst_element}). Indeed, the groundstate obviously
provides the leading contribution in the r.h.s. of
(\ref{scrho_dst_element}). The factorization of position and
temperature dependences in (\ref{low_temp_assympt}) does not come
out easily from FK formula~(\ref{FKd}), where both position and
temperature are coupled in an absolutely non-trivial way. By the
way, notice that spectral representation~(\ref{scrho_dst_element})
is not well suited at intermediate or high temperatures, because one
has then to sum a large number of contributions associated with
various eigenstates (in particular, classical behaviour
(\ref{classical}) cannot be straightforwardly recovered by using
(\ref{scrho_dst_element})).

\bigskip

\noindent When temperature becomes too small, previous computational methods
become more difficult to handle accurately. Even if some algorithmic
tricks \cite{KrauCommunication} may significantly reduce
calculation-times within a satisfactory accuracy, they do not provide a
better understanding of low-temperature behaviours.

\section{Box formula}\label{rewritingFeynamanKac}

\subsection{The central observations}\label{rempathext}

\noindent Let us consider a symmetric confining potential $V(z)$
with a minimum located at $z=0$. On the one hand, low-temperature
behaviour (\ref{low_temp_assympt}) of the corresponding density
matrix is mainly determined by the local shape of $V(z)$ over finite
length scale $a_0$, which characterizes the spatial extension of the
groundstate wave function $\phi_0(x)$. On the other hand, for
typical paths with size of order $\lambda_D$, time-average potential
$\int_0^{1} \mathrm{d} s\, V \left(x + \lambda_D \xi(s) \right)$
becomes, roughly speaking, of order $\int_0^{\lambda_D} \mathrm{d}
z\, V \left( z \right)/\lambda_D$ when $\lambda_D$ is sufficiently
large. At low temperatures, their contribution to the r.h.s. of
(\ref{FKd}) is then controlled by the large-distance behaviour of
$V(z)$. Thus, low-temperature behaviour (\ref{low_temp_assympt}) is
not provided by typical paths. That argument can be implemented
through semi-quantitative estimations for specific cases. If $V(z)$
diverges as $|z|^n$ ($n>0$), contributions of typical paths to
(\ref{FKd}) behave (discarding multiplicative powers of $\beta$) as
$\exp{\left[-c(\beta)^{1+n/2} \right]}$ with some positive constant
$c$: they are exponentially smaller than leading Boltzmann factor
$\exp{\left[-\beta E_0\right]}$.

\bigskip

\noindent The previous analysis suggests that, at low temperatures,
leading contributions to the r.h.s. of (\ref{FKd}) arise from paths with
a spatial extension of order $a_0$, i.e. from quite small
Brownian bridges with size $|\xi(s)|$ of order $a_0/\lambda_D$.
For such paths, time-average potential
$\int_0^{1} \mathrm{d} s\, V \left(x + \lambda_D \xi(s) \right)$ is
of order $V(a_0)$, so the corresponding Boltzmann factor indeed is
of order $\exp{\left[-\beta E_0\right]}$. Notice that,
paths with very different shapes give raise to similar contributions,
since the Wiener weights of the associated Brownian bridges remain of
order (roughly speaking)
$\exp (-\int_0^{1} \mathrm{d} s (\dot{\xi}(s))^2/2) \sim
\exp (-a_0^2/\lambda_D^2) \sim 1$. Consequently, when the
temperature decreases, only a very tiny subset of paths gives a relevant
contribution to the r.h.s. of (\ref{FKd}). In other words,
important paths are not any more typical but, on the contrary, they
can be viewed as large deviations. This explains why direct numerical
evaluations of (\ref{FKd}) become rather difficult:
the subset of important paths remains, in some sense, hidden in the
entire configurational space.

\bigskip

\noindent In order to extract from (\ref{FKd}) the relevant contributions
at low temperatures, it is tempting to collect all paths with the same finite
spatial extension, and then to sum over all possible extensions. That
procedure is not easy to carry out directly in the
r.h.s. of (\ref{FKd}), within a suitable partition of
functional integration space over Brownian bridges. As described below, it
is more convenient to transform first the density matrix
within the operator representation, by introducing an auxiliary
Hamiltonian which confines the particle inside a box with size $\ell$.

\subsection{The auxiliary Hamiltonian approach}

\noindent Let us introduce the auxiliary Hamiltonian
\begin{eqnarray}\label{intermediatehamiltonian}
H_{\ell}&=&H^0+V+V_\ell,
\end{eqnarray}
where $ H^0$ denotes the kinetic Hamiltonian.
The additional potential $V_{\ell}$ is defined by
\begin{eqnarray}
V_\ell(z)&=&V_0 \left[ 1-\Theta
\left(z+\ell\right)+\Theta \left(z-l\right)
\right],
\end{eqnarray}
where $\Theta$ is the Heavyside function, while $V_0$ denotes
barrier height ($V_0 > 0$). That potential tends to confine the
particle inside a box with extension $\ell$.
Expression~(\ref{intermediatehamiltonian}) reduces to Hamiltonian
(\ref{hamiltonian}) when $\ell$ goes to infinity. In that limit, for
$x$ kept fixed, diagonal part $\langle x |\rho_{\ell}| x \rangle$ of
density matrix $\rho_{\ell}=\exp\left[-\beta H_{\ell} \right]$ goes
to $\langle x |\rho | x \rangle$.

\bigskip

\noindent The identity,
\begin{eqnarray}\label{formulaaciteror}
\langle x |\rho | x \rangle &=& \langle x |\rho_{L}| x \rangle+
\int_{L}^{\infty} \mathrm{d} \ell\  \langle x | \partial_{\ell} \left[ \rho_{\ell} \right]| x \rangle
\end{eqnarray}
which is valid for any reference extension $L$ and for any height
$V_0$ of the confining potential is quite useful for our purpose. We set $L=|x|$, and we consider an infinitely high
potential barrier $V_0$.  Under that limit, the contact density
goes to zero,
\begin{equation}
\lim_{V_0\rightarrow+\infty}\langle x | \rho_{|x|} | x \rangle= 0,
\end{equation}
because the probability for the particle to stay on the boundary vanishes
for infinitely high walls. Thus, identity
(\ref{formulaaciteror}) is rewritten as
\begin{eqnarray}\label{formulaaciterorbis}
\langle x |\rho | x \rangle &=& \lim_{V_0 \to \infty}
\int_{|x|}^{\infty} \mathrm{d} \ell\ \langle x | \partial_{\ell} \left[ \rho_{\ell} \right]| x \rangle.
\end{eqnarray}
The right hand side can be transformed by applying
Dyson formula
\begin{equation}
\partial_{\ell} \left[ e^{- A(\ell)}\right] =
-\int_0^{1}\mathrm{d} s \,
e^{-\left[1-s \right] A(\ell)}\,
\partial_{\ell} \left[A(\ell)\right]\,
e^{-s A (\ell)} ,  \label{formuledyson}
\end{equation}
valid for any operator$A(\ell)$ which depends on $\ell$. This provides
\begin{eqnarray}
\partial_{\ell} \left[ \rho_{\ell} \right] &=&
-\beta \int_0^{1}\mathrm{d} s \int_{-\infty }^{+\infty }
\mathrm{d} z \ e^{-\beta (1-s) H_{\ell} }
\left| z \rangle \langle z \right|
\partial_{\ell} \left[H_{\ell}\right]
e^{-\beta s H_{\ell}}\\
&=&\beta V_0 \int_0^{1}\mathrm{d} s
\int_{-\infty }^{+\infty }\mathrm{d} z \ e^{-\beta (1-s) H_{\ell}}
| z \rangle
\left[\delta(z-\ell)+\delta(z+\ell)
\right] \langle z |e^{-\beta s H_{\ell}}\\
&=&\beta V_0 \int_0^{1}\mathrm{d} s
\left(e^{-\beta (1-s) H_{\ell}}|\ell \rangle
\langle \ell |
e^{-\beta s H_{\ell}}
+e^{-\beta (1-s) H_{\ell}} |-\ell \rangle \langle-\ell |
e^{-\beta s H_{\ell}}\right),
\end{eqnarray}
from which we infer
\begin{eqnarray}\label{2B1}
\rho(x,x,\beta)&=& \lim_{V_0 \to \infty} \beta V_0 \int_{|x|}^{\infty} \mathrm{d} \ell
\int_0^1 \mathrm{d} s \left[ \rho_{\ell} \left(x,\ell,\beta [1-s] \right)
\rho_{\ell} \left(\ell,x,\beta s \right)+\rho_{\ell} \left(x,-\ell,\beta [1-s] \right) \rho_{\ell} \left(-\ell,x,\beta s \right)\right].
\end{eqnarray}
That expression must be considered with some caution because when $V_0$ diverges the different integrands
vanishes. Therefore, additional work is necessary to control 
limit $V_0\to \infty$.

\bigskip

\noindent For that purpose, we rewrite the integrands
in terms of constrained density matrix $\rho_{\ell}^0$  of the {\em free} particle submitted to confining potential $V_{\ell}$.
According to the obvious equality
\begin{eqnarray}
\rho_{\ell}\left(x_i,x_f,\beta\right)&=&
\rho_{\ell}^0\left(x_i,x_f,\beta\right)\frac{\rho_{\ell}\left(x_i,x_f,\beta\right)}
{\rho_{\ell}^0\left(x_i,x_f,\beta\right)},
\end{eqnarray}
and since $\rho_{\ell}/\rho_{\ell}^0$ remains finite when $V_0 \to \infty$, we define
\begin{eqnarray}
g^{\pm}(x,\ell,s,\beta)&=& \lim_{V_0 \to \infty}
(2\pi)^{1/2} \lambda_D (\beta V_0)\rho_{\ell}^0\left(x,\pm\ell,\beta s \right)
\rho_{\ell}^0\left(\pm\ell,x,\beta (1-s) \right).
\end{eqnarray}
After using Feynman-Kac formula for both $\rho_{\ell}$ and $\rho_{\ell}^0$
in ratio $\rho_{\ell}/\rho_{\ell}^0$, we transform (\ref{2B1}) into
\begin{eqnarray}
\rho(x,x,\beta) \!\! &=&\!\!\!\!
\int_{|x|}^{+\infty} \!\! \mathrm{d} \ell
\int_{0}^1 \!\! \mathrm{d} s\
\frac{g^{-}(x,\ell,s,\beta)}{\sqrt{2\pi}\lambda_D}
\left\langle \, \exp{\left[-\beta s \int_0^1 \mathrm{d} u\
V\left( z(u) \right)\right]}\, \right \rangle_{\displaystyle \Omega^-_s}
\!\! \left\langle \, \exp{\left[-\beta(1-s) \int_0^1 \mathrm{d} u\ V
\left( z(u) \right)\right]}\, \right \rangle_{\displaystyle \Omega^-_{1-s}} \nonumber \\
&+&\!\!\!\!
\int_{|x|}^{+\infty} \!\! \mathrm{d} \ell
\int_{0}^1 \!\! \mathrm{d} s\
\frac{g^{+}(x,\ell,s,\beta)}{\sqrt{2\pi} \lambda_D}
\left\langle \,\exp{\left[-\beta s \int_0^1 \mathrm{d} u\
V\left(z(u)\right)\right]}\, \right \rangle_{\displaystyle \Omega^+_s}
\!\! \left\langle \,\exp{\left[-\beta(1-s) \int_0^1 \mathrm{d} u\ V
\left( z(u) \right)\right]}\, \right \rangle_{\displaystyle \Omega^+_{1-s}} .\label{2B2}
\end{eqnarray}
Notation
$\langle \, . \,  \rangle_{\omega}$ denotes an average over a constrained Brownian process
which belongs to a set $\omega$. Paths $z(u)$ are expressed in
terms of Brownian bridges according to
$z(u)=x(1-u)\pm u \ell + \sqrt{s} \lambda_D \xi(u)$ for $\Omega^{\pm}_s$.
Moreover those paths must stay inside the box $[-\ell,+\ell]$, so
that constraint defines the corresponding sets $\Omega^{\pm}_s$.
The statistical weight of a path in such a constrained average is its associated Wiener measure $\mathcal{D}_W(\xi)$. For the sake of notational 
convenience, we do not explicitly write the dependences on both $x$ and 
$\ell$ of $\Omega^{\pm}_s$.

\bigskip

\begin{figure}[ht!]
\begin{center}
\resizebox{0.6\textwidth}{!}{\includegraphics{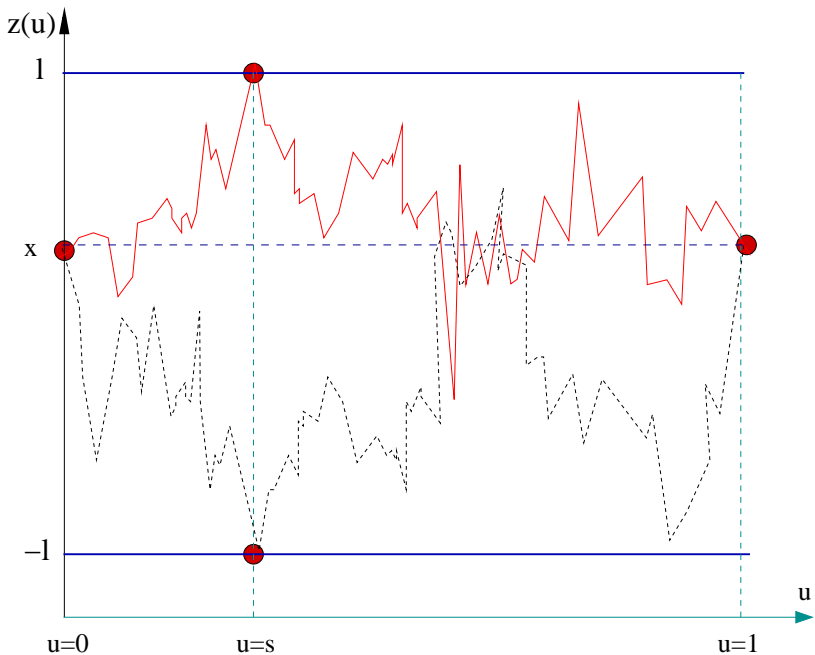}}
\caption{Typical paths which start and end at position
$x$. Solid and dashed lines represent paths which belong to $\Omega^+$
and to $\Omega^-$ respectively. }\label{defbox}
\end{center}
\end{figure}

\noindent The physical interpretation of
functions $g\pm$ clearly emerges from (\ref{2B2}), if we
specify that general formula to the particular case $V(z)=0$. Such functions
are the (normalized) statistical weights of the constrained sets
$\Omega^{\pm}=\Omega^{\pm}_s \bigcup \Omega^{\pm}_{1-s}$, {\it i.e.} the sum of statistical weights of all paths touching the boundaries
of the box  at time $s$ (see  Figure \ref{defbox}). Therefore, in
the following, we set $g^{\pm}(x,\ell,s,\beta)=g(\Omega^{\pm})$.
Analytical calculations of those weights are performed in Appendix A,
by using the spectral representation of $\rho_{\ell}^0$.

\bigskip

\noindent The first step of our rewriting of Feynman-Kac
representation, is achieved through formula~(\ref{2B2}). Paths are
indeed collected together according to their extension $\pm \ell$.
Notice that the touching time, $s$, is also crucial for defining the
corresponding proper partition of the genuine integration space over
all unconstrained paths.

\subsection{Introduction of averages over occupation times}

\noindent In a second step, we introduce the so-called
{\em occupation time}, defined for each given path $z(u)$ in
$\Omega^{\pm}_s$ by
\begin{equation}
\theta_{z}(x')=\int_0^1 \mathrm{d} u \,\delta \left( x'-z(u) \right).
\end{equation}
The quantity $\theta _z(x') \mathrm{d} x'$ is the total time passed
in a neighborhood~$\mathrm{d}x'$ of position~$x'$ by the particle
when it follows Brownian path $z(u)$. Of course, the total time passed in
the whole box is always equal to $1$, {\it i.e}.
\begin{equation}
\int_{-\ell}^{+\ell} \mathrm{d} x' \, \theta_{z}(x') = 1 .
\label{NormaOcu}
\end{equation}
Time-averaged potential along
path $z(u)$ is then expressed in terms of occupation time \textsl{via}
the obvious identity
\begin{equation}
\int_0^1 \mathrm{d} u\, V \left(z(u)\right)=\int_{-\ell}^{+\ell}
\mathrm{d} x' \, \theta_{z}(x')\, V \left(x' \right),
\label{AveOcu}
\end{equation}
valid for any path $z(u)$.

\bigskip

\begin{figure}[ht]
\begin{center}
\resizebox{0.4\textwidth}{!}{\includegraphics{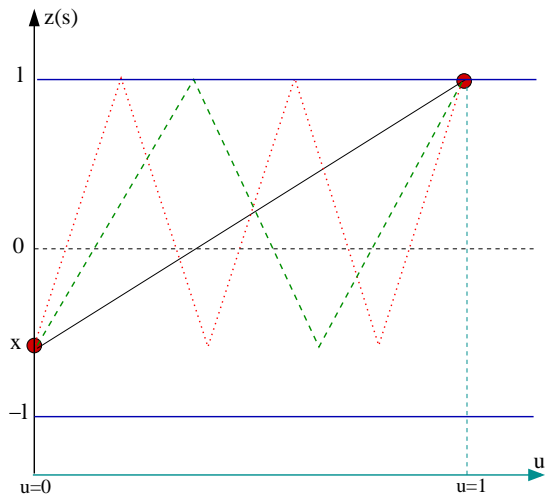}}
\caption{Three different paths which start at $x$ and end
at $+\ell$, while the corresponding occupation times
are identical.}
\label{diffpathsametheta}
\end{center}
\end{figure}

\noindent According to identity~(\ref{AveOcu}), Boltzmann factors involved in
averages
\begin{equation}
\left\langle \, \exp{\left[-\beta s \int_0^1 \mathrm{d} u\
V\left( z(u) \right)\right]}\, \right \rangle_{\displaystyle \Omega^{\pm}_s},
\label{AveBol}
\end{equation}
only depend on the occupation time $\theta_z(x')$. As illustrated in
Fig.~\ref{diffpathsametheta}, various different paths may provide
the same occupation time.
As quoted in Section~\ref{rempathext}, their statistical
weights are close together,
so their contributions to average~(\ref{AveBol}) are almost identical.
Thus, it is now tempting to collect all paths which provide the same
occupation time $\theta(x')$, \textsl{via} the introduction of
the corresponding density measure
$\mathcal{D}_{\Omega^{\pm}_s}[\theta]$. After
expressing averages~(\ref{AveBol}) over constrained paths, as averages over
occupation times with (normalized) measure $\mathcal{D}_{\Omega^{\pm}_s}[\theta]$,
we eventually obtain the box formula
\begin{eqnarray}
\rho(x,x,\beta)\!\!\!\! &&\!\!\!\! =\!\!
\int_{|x|}^{+\infty} \!\!\!\!\!\!\!\! \mathrm{d} \ell \!\!
\int_{0}^1 \!\!\!\! \mathrm{d} s \,
\frac{g(\Omega^{-})}{\sqrt{2\pi}\lambda_D}\!\!
\int \!\! \mathcal{D}_{\Omega^{-}_s}[\theta] \exp\!\!\left[-\beta s \!\! \int_{-\ell}^{+\ell}\!\!\!\! \mathrm{d} x'
\theta(x') V\left( x' \right)\right] \!\!
\int \!\! \mathcal{D}_{\Omega^{-}_{1-s}}[\theta]
\exp\!\!\left[-\beta (1-s) \!\! \int_{-\ell}^{+\ell}\!\!\!\! \mathrm{d} x'
\theta(x') V\left( x' \right)\right]\nonumber \\
&+&\!\!\!\!
\int_{|x|}^{+\infty} \!\!\!\!\!\!\!\! \mathrm{d} \ell
\int_{0}^1 \!\!\!\! \mathrm{d} s\,
\frac{g(\Omega^{+})}{\sqrt{2\pi} \lambda_D}\!\!
\int\!\! \mathcal{D}_{\Omega^{+}_s}[\theta] \exp\!\!\left[-\beta s \!\! \int_{-\ell}^{+\ell}\!\!\!\! \mathrm{d} x'
\theta(x') V\left( x' \right)\right] \!\!
\int \!\!  \mathcal{D}_{\Omega^{+}_{1-s}}[\theta]
\exp\!\!\left[-\beta (1-s) \!\! \int_{-\ell}^{+\ell}\!\!\!\! \mathrm{d} x'
\theta(x') V\left( x' \right)\right].
\label{2B3}
\end{eqnarray}
If statistical weights $g(\Omega^{\pm})$ are analytically known
(see Appendix A), explicit expressions for density measures $\mathcal{D}_{\Omega^{\pm}_s}[\theta]$
are not available. Such probability densities result from the summation
of Wiener measures over all Brownian paths inside $\Omega^{\pm}_s$ which
provide the same occupation time. That procedure is quite difficult
to handle in closed analytical forms, and only the moments of
$\mathcal{D}_{\Omega^{\pm}_s}[\theta]$ can be computed explicitly.
Nevertheless, box formula~(\ref{2B3}) turns out to be quite useful
for various purposes, as suggested by the following simple
comments and arguments.

\bigskip

\noindent Contrary to the case of the genuine Feynman-Kac representation,
leading contributions at low temperature merely emerge from
box formula~(\ref{2B3}). Indeed, for $x$ of order $a_0$, boxes with size
$\ell$ of order a few $a_0$ do provide contributions of order
$\exp (- \beta E_0)$, because weight factors $g(\Omega^{\pm})$
are of order $\exp (-\beta \hbar^2/(2m \ell^2))$
for $\ell \ll \lambda_D$ (see Appendix A), while products of averages
of Boltzmann factors over occupation times are of order
$\exp (- \beta \int_0^{\ell} \mathrm{d} x' V\left( x' \right)/\ell)$.
That rough analysis will be implemented in a forthcoming paper,
where we show more precisely how low temperature
behaviour (\ref{low_temp_assympt}) arises from scaling properties
of distributions $\mathcal{D}_{\Omega^{\pm}_s}[\theta]$ in the
regime $\ell \ll \lambda_D$.

\bigskip

\begin{figure}[ht]
\begin{center}
\resizebox{0.7\textwidth}{!}{\includegraphics{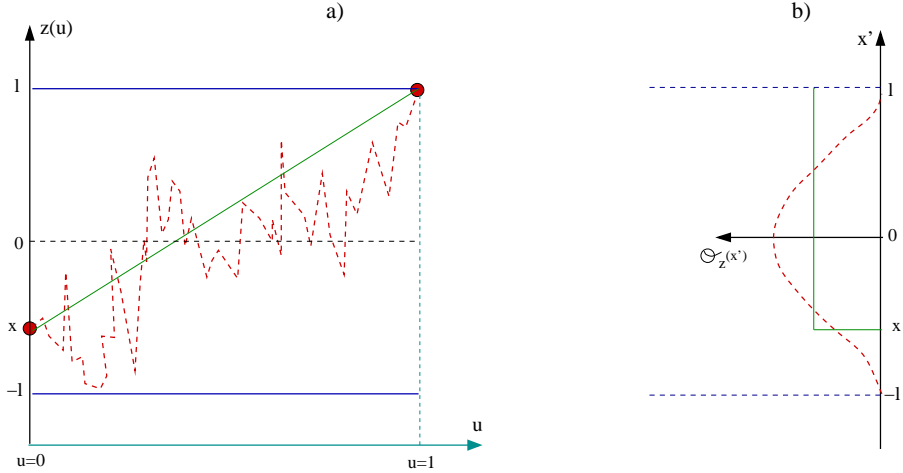}}
\caption{Panel (a) presents two different paths  which start at position
  $x$ at time $u=0$ and end at $\ell$ for time
  $u=1$. Panel (b) shows their corresponding occupation times
  $\theta_z(x')$. }\label{diffpathettheta}
\end{center}
\end{figure}

\noindent If Brownian paths are quite noisy, the corresponding occupation
times may have rather regular shapes. This is illustrated in
Fig.~\ref{diffpathettheta}$a$, which shows a regular path on the one hand, an
a very jagged one on the other hand: both paths provide
occupation times with regular shapes displayed in
Fig.~\ref{diffpathettheta}$b$ (notice that the corresponding contributions
to Boltzmann factors are of course different).
That observation leads to a first type of approximations
based on simple modelizations of distributions
$\mathcal{D}_{\Omega^{\pm}_s}[\theta]$ within
restricted sets of elementary functions which represent the various
occupation times (they will be described elsewhere). A second type
of approximations is based on the truncation of cumulant expansions,
where key ingredients do exhibit regular behaviours with respect to
spatial positions. One of them is presented further in
Section~\ref{constrainedaverages}.

\bigskip

\noindent Eventually, notice that box formula~(\ref{2B3}) can be immediately
extended to any potential $V(x)$, not necessarily confining.
Similar formula can be derived
in higher dimensions or for systems with more than one particle.
Those various extensions may include different box shapes or
different confining potentials $V_0(x)$.

\section{Ergodic approximation}
\label{constrainedaverages}

\subsection{Truncation of the cumulant expansion}

\noindent Formally, averages over distributions $\mathcal{D}_{\Omega^{\pm}_s}[\theta]$ involved in the
r.h.s. of box formula~(\ref{2B3}) can be represented by their
infinite cumulant expansions. A natural approximation
consists in truncating that expansion up to its first term, {\it i.e}.
we replace the averages
\begin{eqnarray}
\int \mathcal{D}_{\Omega^{\pm}_s}[\theta] \exp\!\!\left[-\beta s \!\! \int_{-\ell}^{+\ell}\!\!\!\! \mathrm{d} x'
\theta(x') V\left( x' \right)\right]
\end{eqnarray}
by
\begin{eqnarray}\label{cum1}
\exp{\left(-\beta s
\int_{-\ell}^{+\ell} \mathrm{d} x'\,
< \theta (x') >_{\Omega^{\pm}_s}\, V \left( x'\right) \right)} .
\end{eqnarray}
We call that lowest order approximation {\em ergodic}, since it would
be exact if paths were experiencing all parts of the potential with a
probability independent of time. It amounts to replace (imaginary) time
averages of the potential by spatial averages with
a measure defined by mean occupation-time
\begin{equation}
< \theta (x') >_{\Omega^{\pm}_s} =
\int \mathcal{D}_{\Omega^{\pm}_s}[\theta] \, \theta (x').
\label{MeanOccu}
\end{equation}
Inserting (\ref{cum1}) into (\ref{2B3}), we obtain the
subsequent ergodic approximation for density matrix,
\begin{eqnarray}\label{est1_1b}
\rho_{erg}(x,x,\beta) \!\! &=&\!\!\!\!
\int_{|x|}^{+\infty} \!\!\!\!\!\!\!\! \mathrm{d} \ell
\int_{0}^1 \!\! \mathrm{d} s\
\frac{g(\Omega^{-})}{\sqrt{2 \pi}\lambda_D} \exp{\left[-\beta \int_{-\ell}^{+\ell} \mathrm{d} x'
\left(s < \theta (x') >_{\Omega^{-}_s} +
[1-s] < \theta (x') >_{\Omega^{-}_{1-s}}\right)
 V\left( x' \right)\right]} \nonumber \\
&+&\!\!\!\!
\int_{|x|}^{+\infty} \!\!\!\!\!\!\!\! \mathrm{d} \ell
\int_{0}^1 \!\! \mathrm{d} s\
\frac{g(\Omega^{+})}{\sqrt{2 \pi}\lambda_D} \exp{\left[-\beta \int_{-\ell}^{+\ell} \mathrm{d} x'
\left(s < \theta (x') >_{\Omega^{+}_s} +
[1-s] < \theta (x') >_{\Omega^{+}_{1-s}}\right)
 V\left( x' \right)\right]} .
\end{eqnarray}

\subsection{Mean occupation time in terms of Brownian Green functions}

\noindent The average occupation time can be obtained within two different methods: either by using operator
algebra (see Appendix B) or, in a more physical way, by using
Brownian motion properties, as shown below.
First we compute the mean occupation time around position $x'$,
over Brownian paths at inverse temperature $\beta'$ which start from $x_i$ at time $u=0$ and end at
$x_f$ for time $u=1$. Those Brownian paths are
constrained within the interval $[-\ell, +\ell]$, and they define a
set $\omega=\{ z(u)=x_i(1-u)+x_fu+\lambda \, \xi(u) \}$
with $\lambda^2=\beta' \hbar^2/m$.

\bigskip

\noindent Introducing the constrained
probability density $f_{\omega}(z,u)$ to find the Brownian
particle at the position $z$ for time $u$, the
mean occupation time reads
\begin{eqnarray}
\left\langle \theta (x') \right\rangle_{\omega}&=&\int_{0}^{1} \mathrm{d} u\, f_{\omega}(x',u) .\label{mteta_1b}
\end{eqnarray}
The probability density
$f_{\omega}(x',u)$ can be
generated by using the Green function $G_{\omega}(z,t|z_0,t_0)$ of the diffusion
equation, solved with Dirichlet boundary conditions
$G_{\omega}(\pm\ell,t|z_0,t_0)=0$ and initial condition
$G_{\omega}(z,t_0|z_0,t_0)=\delta(z-z_0)$. That Green function reduces to
\begin{equation} \label{green}
G_{\omega}(z,t|z_0,t_0)={1 \over \ell}
\sum_{n=1}^{\infty}  \psi_n(z_0/\ell) \psi_n(z/\ell) \exp{\left[-
(t-t_0) \frac{\lambda^2}{\ell^2}
\frac{\pi^2}{8}  n^2\right]},
\end{equation}
with $\psi_n(z/\ell)=\sin{ [n \pi (z+\ell)/ (2 \ell) ] }$.
Since Brownian motion is a Markov process, the constrained probability
density $f_{\omega}(x',u)$  is then
\begin{eqnarray}
f_{\omega}(x',u)=\frac{\displaystyle G_{\omega}(x',u|x_i,0) \, G_{\omega}(x_f,1-u|x',0)}{G_{\omega}(x_f,1|x_i,0)}.
\end{eqnarray}

\bigskip

\noindent
Now, we focus on the specific case $\omega=\Omega^{-}_s$ ({\it i.e.} $x_f=-\ell$, $x_i=x$ and $\beta'=\beta s$).
According to (\ref{green}), the Green function $G_{\Omega^{-}_s}(x',u|x,0)$ can be rewritten as
\begin{eqnarray}
\displaystyle G_{\Omega^{-}_s}(x',u|x,0)=\frac{1}{\ell} \,
\eta_1\left(\frac{x}{\ell},\frac{x'}{\ell},\frac{\lambda_D\sqrt{s}}{\ell},u\right),
\end{eqnarray}
where $\eta_1$ is a dimensionless function depending on four dimensionless variables.
Similarly, we find
\begin{eqnarray}
\displaystyle \frac{G_{\Omega^{-}_s}(-\ell,1-u|x',0)}{G_{\Omega^{-}_s}(-\ell,1|x,0)}
=\eta_2 \left(\frac{x}{\ell},\frac{x'}{\ell},\frac{\lambda_D\sqrt{s}}{\ell},u\right).
\end{eqnarray}
If we introduce the dimensionless function $\Phi$  defined as
\begin{eqnarray}\label{phiinteg}
\Phi \left(\alpha',\alpha,y\right)=
\int_0^1 \mathrm{d} u \ \eta_1(\alpha,\alpha',y,u) \eta_2(\alpha,\alpha',y,u),
\end{eqnarray}
we obtain
\begin{eqnarray}
\left\langle \theta (x') \right\rangle_{\Omega^{-}_s}&=&
\frac{1}{\ell} \Phi \left(\frac{x}{\ell},\frac{x'}{\ell},\frac{\lambda_D\sqrt{s}}{\ell}\right).
\end{eqnarray}
Furthermore, thanks to the symmetry properties of Green functions,
$G_{\omega}(z,t|z_0,t_0)=G_{\omega}(z_0,t|z,t_0)=G_{\omega}(-z,t|-z_0,t_0)$,
$\left\langle \theta (x') \right\rangle_{\Omega^{+}_s}$ is also given by
\begin{eqnarray}
\left\langle \theta (x') \right\rangle_{\Omega^{+}_s}&=&
\frac{1}{\ell} \Phi \left(\frac{-x}{\ell},\frac{-x'}{\ell},\frac{\lambda_D\sqrt{s}}{\ell}\right).
\end{eqnarray}
Inserting those expressions of mean occupation times for various
sets $\Omega^{\pm}_s$ into (\ref{est1_1b}), the ergodic form of
the density matrix is rewritten as
\begin{eqnarray}
\rho_{erg}(x,x,\beta) \!\! &=&\!\!\!\!\int_{|x|}^{+\infty} \!\!\!\!\!\!\!\! \mathrm{d} \ell
\int_{0}^1 \!\! \mathrm{d} s\,
\frac{g(x,\ell,s,\beta)}{\sqrt{2 \pi}\lambda_D}\exp{\left(-\beta \int_{-1}^{1} \!\!\!\!\mathrm{d} \alpha'
\left[s \Phi \left(\frac{x}{\ell},\alpha',\frac{\pi \lambda_D \sqrt{s}}{2 \ell}\right)+
[1-s] \Phi \left(\frac{x}{\ell},\alpha',\frac{\pi \lambda_D \sqrt{1-s}}{2 \ell}\right)
\right] V\left( \alpha' \ell \right)\right)} \nonumber \\
&+&\!\!\!\!
\int_{|x|}^{+\infty} \!\!\!\!\!\!\!\! \mathrm{d} \ell
\int_{0}^1 \!\! \mathrm{d} s\,
\frac{g(-x,\ell,s,\beta)}{\sqrt{2 \pi}\lambda_D} \exp{\left(-\beta \int_{-1}^{1}\!\!\!\!\mathrm{d} \alpha'
\left[s \Phi \left(\frac{-x}{\ell},\alpha',\frac{\pi \lambda_D \sqrt{s}}{2 \ell}\right)+
[1-s] \Phi \left(\frac{-x}{\ell},\alpha',\frac{\pi \lambda_D \sqrt{1-s}}{2 \ell}\right)
\right] V\left(- \alpha' \ell \right)\right)} \label{est2_1b}
\end{eqnarray}
with $g(x,\ell,s,\beta) = g(\Omega^{-}_s)$. In (\ref{est2_1b}),
all quantities are explicitly known in terms of simple series
involving Gaussian and trigonometric functions. Box weight
$g(x,\ell,s,\beta)$ is computed in Appendix A, while a similar expression is
derived for $\Phi$ in the next subsection (see formula (\ref{Phigene})).
Within the ergodic approximation, we are left with the evaluation of two ordinary integrals, which is of course much easier than a direct evaluation
of the genuine functional integrals.

\subsection{Analytical estimations for  $\Phi$ at low and high temperatures}

\subsubsection{General expression for $\Phi$}

\bigskip

\noindent First, we derive a general expression for $\Phi$, 
valid at any temperature. The spectral expressions of $\eta_1$ and 
$\eta_2$ are
\begin{eqnarray}
\eta_1 \left(\alpha,\alpha',y,u\right)&=&\sum_{n=1}^{\infty}
\psi_n(\alpha)\psi_n(\alpha')\exp{\left[-\frac{ y^2}{2} n^2 u\right]},\label{formalphi1}\\
\eta_2 \left(\alpha,\alpha',y,u\right)&=& \frac{\displaystyle \sum_{n=1}^{\infty}
\psi_n(\alpha') n
\exp{\left[-\frac{y^2}{2}  n^2 [1-u]  \right]}}{\displaystyle \sum_{n=1}^{\infty}
\psi_n(\alpha) n \exp{\left[-\frac{y^2}{2}  n^2 \right]}}\label{formalphi23}.
\end{eqnarray}
Inserting those expressions into formula~(\ref{phiinteg}), we find the general expression
for $\Phi$,
\begin{eqnarray}\label{Phigene0}
\Phi \left(\alpha,\alpha',y\right)=
\frac{\displaystyle \sum_{n=1}^{\infty} \psi_n(\alpha)\psi_n^2(\alpha')n \exp{\left[-\frac{y^2}{2}  n^2\right]}
+\frac{2}{y^2} \sum_{n\neq k, 1}^{\infty} \psi_n(\alpha)\psi_n(\alpha') \psi_k(\alpha') k
\frac{\exp{\left[-\frac{y^2}{2} n^2 \right]}-\exp{\left[-\frac{y^2}{2} k^2 \right]}}{k^2-n^2}}{\displaystyle \sum_{n=1}^{\infty}
\psi_n(\alpha) n \exp{\left[-\frac{y^2}{2}n^2 \right]}}.
\end{eqnarray}
Normalization condition $\int_{-1}^{+1} \Phi(\alpha,\alpha',y)=1$, 
which follows from (\ref{NormaOcu}), is indeed satisfied by 
(\ref{Phigene0}) thanks to the orthogonality of the $\psi_n$'s, {\it i.e} 
$\int_{-1}^{+1} \mathrm{d} x \,  \psi_n(x)\psi_k(x)= \delta_{k,n}$.
Within simple algebra, that last expression can be rewritten as a simple sum~\cite{gradshteyn}
\begin{eqnarray}\label{Phigene}
\Phi \left(\alpha,\alpha',y\right)=
\frac{\displaystyle \sum_{n=1}^{\infty} \exp{\left[-\frac{y^2}{2}n^2\right]}\psi_n(\alpha') \left[ n \psi_n(\alpha)\psi_n(\alpha')
+\frac{2}{y^2} A_n(\alpha,\alpha') \right]}{\displaystyle \sum_{n=1}^{\infty}
\psi_n(\alpha) n \exp{\left[-\frac{y^2}{2} n^2 \right]}},
\end{eqnarray}
with
\begin{eqnarray}\label{An}
A_n(\alpha,\alpha')=\frac{\pi}{8}\left[ (1-\alpha-2\alpha')\sin\left[n \frac{\pi}{2}(\alpha+\alpha'+2)\right] +\left(1+\alpha-2
\alpha'-2\frac{\alpha-\alpha'}{|\alpha-\alpha'|}\right)\sin\left[n
\frac{\pi}{2}(\alpha-\alpha')\right]\right].
\end{eqnarray}

\bigskip

\noindent In formula~(\ref{est2_1b}), third variable $y$ of the function
$\Phi$ is proportional to $\lambda_D/\ell$.  Thus
the low and high temperature regimes correspond to the limits
$y\to \infty$ and $y\to 0$ respectively. In the next subsections, we
derive asymptotic formulas for $\Phi$ in those two limits.

\subsubsection{Low temperature regime}

\begin{figure}[ht]
\begin{center}
\psfrag{x}{\Large{$\alpha'$}} \psfrag{h}{\Large{$\Phi$}}
\resizebox{0.45\textwidth}{!}{\includegraphics{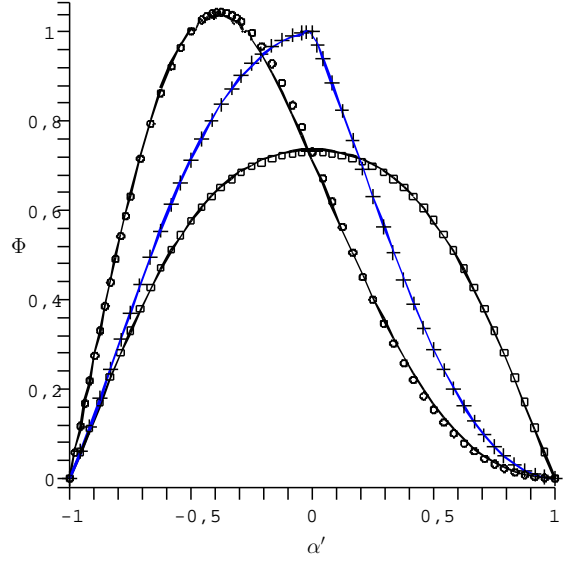}}
\caption{Dimensionless function $\Phi(\alpha,\alpha',2) $ for
 $\alpha=-0.99$ (circles), $\alpha=0$ (plus signs) and  $\alpha=0.99$ (squares).
Solid lines follow from the numerical calculation of expression (\ref{Phigene}), while
points correspond to asymptotic formula (\ref{Phiyassympyg}).}\label{figygrand1}
\end{center}
\end{figure}

\noindent The asymptotic form of $\Phi$ when $y\to\infty$ limit
is obtained by keeping only terms $n=1$ 
in formula~(\ref{Phigene}). This leads to
\begin{eqnarray}\label{Phiyinfty1}
\Phi \left(\alpha,\alpha',y\right)&=&\psi_1^2(\alpha') \left(1+\frac{2}{y^2}\frac{A_1(\alpha,\alpha')}{\psi_1(\alpha)\psi_1(\alpha')}\right)
+O\left(\exp{[-y^2 /2]}\right) \\
&=&\!\!\! \cos^2{\left( \frac{\pi\alpha'}{2}\right)} \left[1+
\frac{\pi}{4 y^2} (\alpha'-1)\tan{\left(\frac{\pi}{2} \alpha'\right)} \right] \nonumber \\
&+&\!\!\! \cos^2{\left( \frac{\pi\alpha'}{2}\right)} \left[
(M_{\alpha,\alpha'}-1)\tan{\left(\frac{\pi}{2} M_{\alpha,\alpha'}\right)} +(1+m_{\alpha,\alpha'})\tan{\left(\frac{\pi}{2}
m_{\alpha,\alpha'}\right)}
\right] \frac{\pi}{4 y^2}+O\left(\exp{[-y^2 /2]}\right),\label{Phiyassympyg}
\end{eqnarray}
where $m_{\alpha,\alpha'} =\min(\alpha,\alpha')$ and $M_{\alpha,\alpha'} =\max(\alpha,\alpha')$.
Notice that the leading the term is normalized to unity (in other words, the leading term satisfies the normalization condition of $\Phi$).
Fig.~\ref{figygrand1} shows the comparison between asymptotic expression (\ref{Phiyassympyg}) and the exact formula (\ref{Phigene}) evaluated numerically.
Asymptotic formula is really
accurate for any $\alpha$, even for $y$ close to unity.

\subsubsection{High temperature regime}

\begin{figure}[ht]
\begin{center}
\resizebox{0.45\textwidth}{!}{\includegraphics{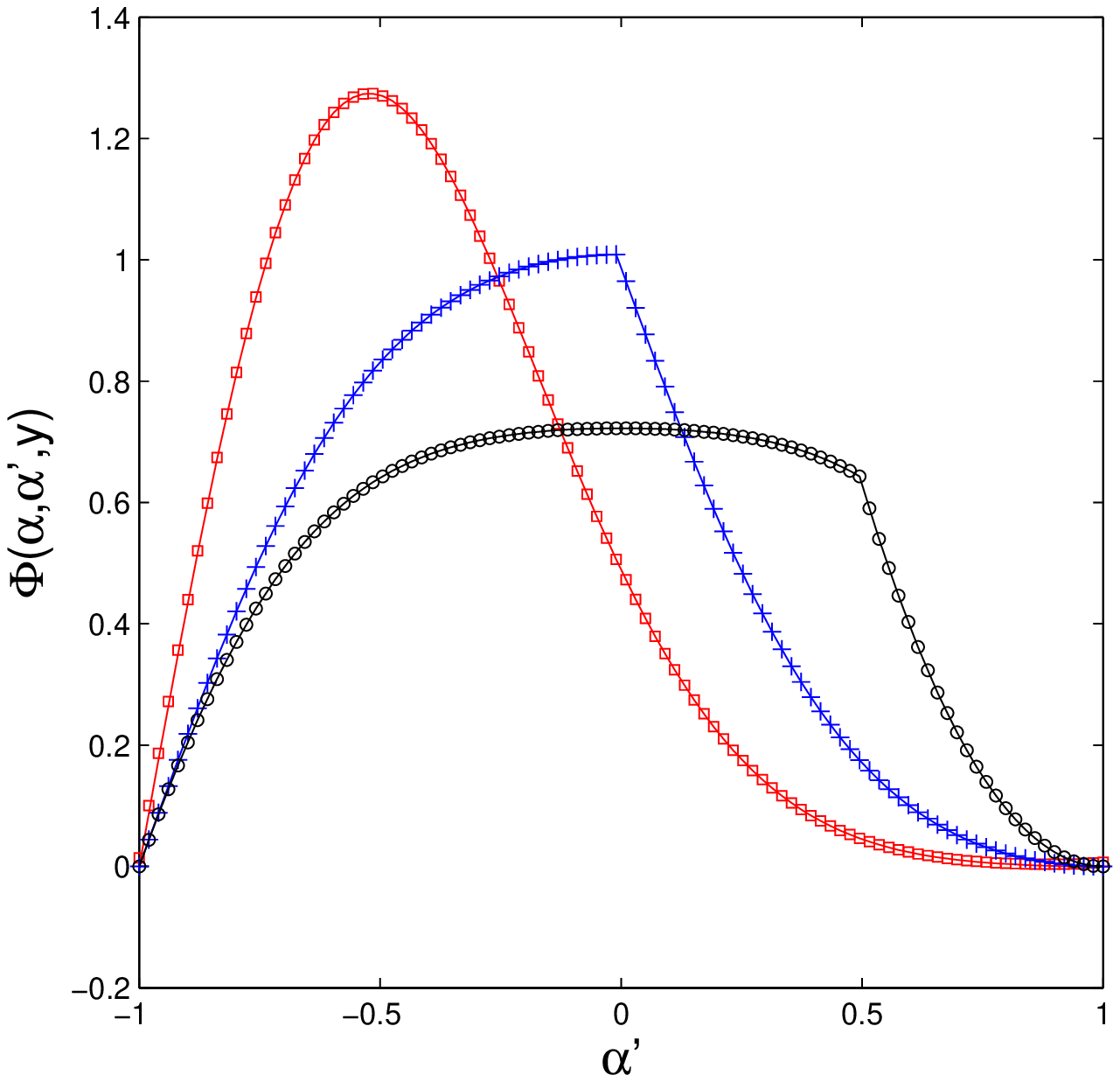}}
\resizebox{0.45\textwidth}{!}{\includegraphics{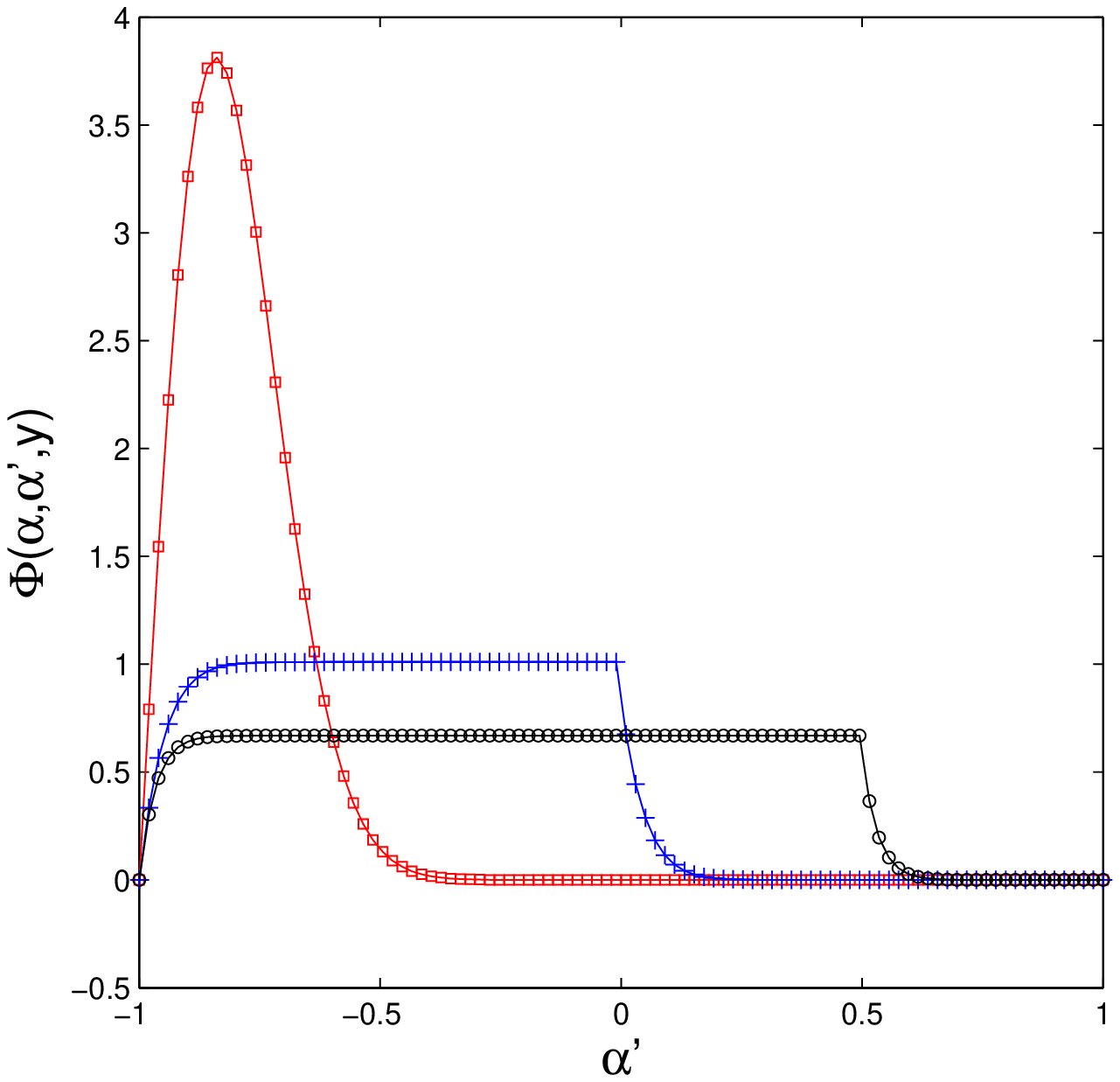}}
\caption{Dimensionless function $\Phi(\alpha,\alpha',1.5)$ (left panel) and $\Phi(\alpha,\alpha',0.5) $ (right panel) for $\alpha=-0.99$ (square) , $\alpha=0$ (cross)
and  $\alpha=0.5$ (circle).
The solid lines follow from numerical calculations of formulas
  (\ref{Phiht1})-(\ref{Phiht2}), while symbols correspond to small-$y$ asymptotic formula.}\label{figypetit}
\end{center}
\end{figure}

In order to obtain the small-$y$ behaviour of $\Phi$, the Poisson transform (\ref{poisson}) is applied to (\ref{Phigene}).
We find
\begin{eqnarray}\label{Phiht1}
\Phi(\alpha,\alpha',y)&=&\frac{1}{2}+\frac{1}{2}\frac{\displaystyle(1-\alpha)+(1-\alpha)
\sum_{n \neq 0}\exp{\left[-\frac{\pi^2}{y^2} n (2n-\alpha-1) \right]}}
{\displaystyle(1+\alpha)+\sum_{n \neq 0}(1+\alpha+4n) \exp{\left[-\frac{\pi^2}{y^2} n (1+\alpha+2n) \right]}} \nonumber \\
&+&\frac{\displaystyle \sum_{n \neq 0}n\exp{\left[-\frac{\pi^2}{2 y^2}
(\alpha'+\alpha-2n)(\alpha'-1-2n) \right]}-\sum_{n \neq 0}n\exp{\left[-\frac{\pi^2}{2 y^2}
(\alpha'-\alpha-2n)(1+\alpha'-2n) \right]}}{\displaystyle(1+\alpha)
+\sum_{n \neq 0}(1+\alpha+4n) \exp{\left[-\frac{\pi^2}{y^2} n (1+\alpha+2n) \right]}}
\end{eqnarray}
for $\alpha'<\alpha$, while
\begin{eqnarray}\label{Phiht2}
\Phi(\alpha,\alpha',y)&=&\frac{1}{2}-\frac{1}{2}\frac{\displaystyle(1+\alpha)-2\exp{\left[-\frac{\pi^2}{2 y^2}
(\alpha'-\alpha)(1+\alpha') \right]}+(1+\alpha)\sum_{n \neq 0}
\exp{\left[-\frac{\pi^2}{y^2} n (2n-\alpha-1) \right]}}
{\displaystyle(1+\alpha)+\sum_{n \neq 0}(1+\alpha+4n) \exp{\left[-\frac{\pi^2}{y^2} n (1+\alpha+2n) \right]}} \nonumber \\
&+&\frac{\displaystyle \sum_{n \neq 0}n\exp{\left[-\frac{\pi^2}{2 y^2}
(\alpha'+\alpha-2n)(\alpha'-1-2n) \right]}+\sum_{n \neq 0}(1-n)
\exp{\left[-\frac{\pi^2}{2 y^2} (\alpha'-\alpha-2n)(1+\alpha'-2n)
\right]}}{\displaystyle(1+\alpha)+\sum_{n \neq 0}(1+\alpha+4n)
\exp{\left[-\frac{\pi^2}{y^2} n (1+\alpha+2n) \right]}}
\end{eqnarray}
for $\alpha'>\alpha$. For $y$ small, numerical estimates of those expressions are obtained by
truncating all sums to terms  $n=\pm 1$.
They are compared with exact formulas (\ref{Phiht1})-(\ref{Phiht2}) in
Fig.~\ref{figypetit} and are quite accurate
for any $\alpha$, and even for $y$ close to unity.
Let us notice that, in the very high temperature regime and for almost all values of $\alpha$,
$\Phi$ is close to $\Theta(\alpha-\alpha')/(1+\alpha)$.

\section{Applications of the ergodic approximation for confining potentials}\label{applications}

\noindent In the remainder of the paper, we will restrict ourselves
to a symmetric confining potential.  Let us focus at first on the
following question: does the ergodic approximation lead to
reliable estimates in the low temperature regime ? Fully
analytical results  will be presented for the simplest potentials,
namely the harmonic potential and the infinite square well.
For more complicated potentials, we combine both
analytical and numerical methods.

\subsection{Low temperature regime}\label{app_ana1}

\subsubsection{General asymptotic formula}

\noindent Inserting the asymptotic expressions (\ref{Phiyassympyg}) and
(\ref{gbt}) for $\Phi$ and $g$ (respectively) into formula (\ref{est2_1b}),
we find the low-temperature behaviour of the ergodic
density matrix,
\begin{eqnarray}\label{geneconfin1}
\rho_{erg}(x,x,\beta) \!\!\!&\stackrel{\beta \to \infty}{\sim}&\!\!\!
\int_{|x|}^{+\infty} \mathrm{d} \ell
\frac{\pi^2 \lambda_D^2}{8 \ell^4}
\sin^2{\left[\frac{\pi(x+\ell)}{2 \ell}\right]} \exp{\left[-\beta E(\ell)\right]}
\left( \exp{\left[-\varphi^+(\ell,x/\ell)\right]}+\exp{\left[-\varphi^-(\ell,-x/\ell)\right]} \right),
\end{eqnarray}
with
\begin{eqnarray}
E(\ell)&=&\frac{\pi^2 \hbar^2}{8 m \ell^2}
+\int_{-1}^{+1} \mathrm{d} \alpha' \cos^2\left(\frac{\pi\alpha'}{2}\right) V(\alpha' \ell),\label{E1gene}\\
\varphi^{\pm}(\ell,\alpha)&=&\frac{2 m \ell^2}{\pi \hbar^2} \int_{-1}^{+1} \!\!
\mathrm{d} \alpha' \, (\alpha'-1)\tan{\left(\frac{\pi}{2} \alpha'\right)}
\cos^2\left(\frac{\pi\alpha'}{2}\right) V(\pm \alpha' \ell) \nonumber\\
       &+& \frac{2 m \ell^2}{\pi \hbar^2} \int_{-1}^{+1} \!\! \mathrm{d} \, \alpha'  V(\pm \alpha' \ell)
       \cos^2\left(\frac{\pi\alpha'}{2}\right)\left[
(M_{\alpha,\alpha'}-1)\tan{\left(\frac{\pi}{2} M_{\alpha,\alpha'}\right)} +(1+m_{\alpha,\alpha'})\tan{\left(\frac{\pi}{2}
m_{\alpha,\alpha'}\right)}\right] \nonumber \\
      &=&\frac{2 m \ell^2}{\pi \hbar^2}\left[2\int_0^1\mathrm{d}\alpha' V(\alpha' \ell)
\sin(\pi \alpha')\alpha'
-\int_{\alpha}^1 \mathrm{d}\alpha' V(\alpha' \ell) \sin(\pi\alpha')\right]\nonumber \\
&+& \frac{2 m \ell^2}{\pi \hbar^2} \alpha \tan\left(\frac{\pi \alpha}{2}\right)\left[\int_0^1 \mathrm{d}\alpha'V(\alpha' \ell) \cos^2\left(\frac{\pi
\alpha'}{2}\right)-\frac{1}{\alpha} \int_0^{\alpha}\mathrm{d}z \, V(\alpha'\ell) \cos^2\left(\frac{\pi
\alpha'}{2}\right)\right].
\end{eqnarray}
Positivity of $\varphi^{\pm}$ enforces the convergence of the integral
in the r.h.s. of (\ref{geneconfin1}) (contributions from boxes with
large sizes $\ell$ do vanish in an integrable way).
Expression~(\ref{geneconfin1}) can still be simplified in the
low temperature regime ($\beta \rightarrow \infty$), by using
the saddle point method.  The divergence of potential $V(z)$ for $z$ large ensures the existence of a value $\ell_0$ which minimizes the function
$E(\ell)$. We emphasize that $\ell_0$
depends, of course, on the potential, and is attained in
integral~(\ref{geneconfin1}) only for values of $x$ such that
$|x|<\ell_0$. Assuming that the second derivative of $E$ is well
defined at $\ell_0$, we can apply the saddle point method for
$|x|<\ell_0$. This provides the general result
\begin{eqnarray}\label{geneconfin2}
\rho_{erg}(x,x,\beta) \!\!\!&\stackrel{\beta \to \infty}{\sim}&\!\!\!n(x)\, \Gamma(\beta)
\end{eqnarray}
for $|x|<\ell_0$. In (\ref{geneconfin2}), $n(x)$ is the (unnormalized) density
\begin{eqnarray}
n(x)&=& \frac{1}{\ell_0}
\sin^2{\left[\frac{\pi(x+\ell_0)}{2 \ell_0}\right]}\, \left(
\exp{\left[-\varphi^+(\ell_0,x/\ell_0)\right]}
+\exp{\left[-\varphi^-(\ell_0,-x/\ell_0)\right]} \right),\end{eqnarray}
while the temperature dependence is entirely embedded into
\begin{eqnarray} \Gamma(\beta)&=&
\frac{\pi^2 \lambda_D}{8 \ell_0} \sqrt{\frac{2 \pi \hbar^2}{m \ell_0^4 E''(\ell_0)}}\exp{\left[-\beta E(\ell_0)\right]}.
\end{eqnarray}
We stress that temperature and position dependences
are factorized in formula~(\ref{geneconfin2}), like in the
exact low-temperature behaviour (\ref{low_temp_assympt}).
Although the temperature dependence of the
ergodic density matrix is not entirely correct (because of the presence of
factor $\lambda_D$ in $\Gamma(\beta)$), the quantity $E(\ell_0)$ can be
identified as the ergodic groundstate energy: indeed, it
controls the exponential decay of $\rho_{erg}(x,x,\beta)$ when
$\beta \rightarrow \infty$. Below, we show, through several examples,
that $E(\ell_0)$ is a good approximation for the real groundstate energy.

\bigskip

\noindent For $|x| > \ell_0$, the saddle point $\ell_0$
is outside the integration range of (\ref{geneconfin1}). In that case,
$\rho_{erg}(x,x,\beta)$ could be evaluated by expanding the
involved integrand for $\ell$ close to $|x|$. Of course, for a 
given $\beta$, if $x$ becomes sufficiently large, 
$\rho_{erg}(x,x,\beta)$ tends to the classical Boltzmann factor 
(which, by the way, vanishes exponentially fast).

\subsubsection{Infinite square potential}

\noindent First, let us consider
the infinite square well potential
\begin{equation}
V(z)=\left\{ \begin{array}{ll}
-V_0& \mbox{if}\quad z<|\ell_p|  \\
+\infty &  \mbox{if}\quad  z >|\ell_p| \end{array}
  \right .
\end{equation}
which is the simplest symmetric and confining potential. Identity
\begin{eqnarray}\label{somymo}
\int_{-1}^{+1} \mathrm{d}\alpha' \cos^2{\left( \frac{\pi\alpha'}{2}\right)} \left[
(M_{\alpha,\alpha'}-1)\tan{\left(\frac{\pi}{2} M_{\alpha,\alpha'}\right)} +(1+m_{\alpha,\alpha'})\tan{\left(\frac{\pi}{2}
m_{\alpha,\alpha'}\right)}+(\alpha'-1)\tan{\left(\frac{\pi}{2} \alpha'\right)}
\right]=0
\end{eqnarray}
allows us to recast expression~(\ref{geneconfin1}) as
\begin{eqnarray}
\rho_{erg}(x,x,\beta) \!\!\!&\stackrel{\beta \to \infty}{\sim}&\!\!\!\
\theta(\ell_p-|x|) \frac{1}{\ell_p}
\sin^2{\left[\frac{\pi(x+\ell_p)}{2 \ell_p}\right]}\exp{\left[\beta V_0 \right]}
\exp{\left[-\beta \frac{\pi^2 \hbar^2}{8 m \ell_p^2} \right]}.
\end{eqnarray}
Therefore, the ergodic approximation gives the exact
groundstate properties of an infinite square well.
That exact results cannot be retrieved within usual approximations, like the semi-classical one.

\subsubsection{Harmonic potential}

\begin{figure}[ht]
\begin{center}
\psfrag{x}{\Large{$x \left( m \omega / \hbar \right)^{1/2} $}} \psfrag{y}{\large{$\Psi_{(1)}(x)$}}
\resizebox{0.4\textwidth}{!}{\includegraphics{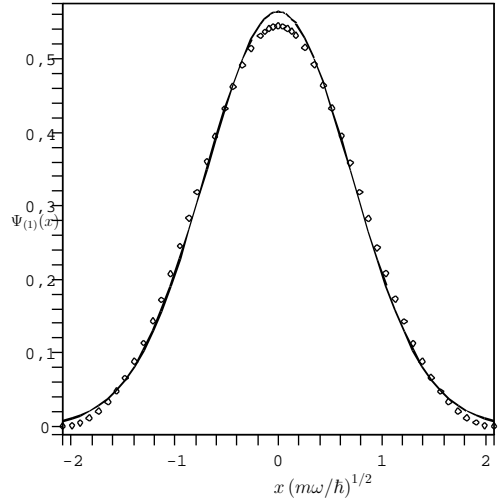}}
\caption{Probability density of the ground state for the harmonic oscillator. The solid line represents
the exact solution while points correspond to the ergodic approximation (\ref{zetaOH}).}
\label{fig_pdf_OH}
\end{center}
\end{figure}

\noindent For the harmonic potential $V(z)=m \omega^2 z^2/2$, we find
\begin{eqnarray}
E(\ell)
=\frac{\pi^2 \hbar^2}{8 m \ell^2}+m \omega^2 \ell^2 \frac{(\pi^2-6)}{6 \pi^2}.
\end{eqnarray}
The minimum of that expression is reached for
\begin{eqnarray}
\ell_0=\left(\frac{\pi^2 \hbar }{2 m \omega}\sqrt{\frac{3}{\pi^2 -6}}\right)^{1/2},
\end{eqnarray}
and it reduces to
\begin{eqnarray}
E(\ell_0)&=&\frac{\hbar \omega}{2}\left[\frac{\pi^2-6}{3}\right]^{1/2} \thickapprox 1.14 \frac{\hbar \omega}{2}.
\end{eqnarray}
Since the second derivative  $E''(\ell_0)= 4 m
\omega^2 ({\pi^2-6})/({3\pi^2})$ is well defined, it is possible to
apply formula~(\ref{geneconfin2}) for $|x| < \ell_0$.
Using
\begin{eqnarray}
\varphi^{\pm}(\ell_0,\pm x/\ell_0)
&=&\frac{1}{4(\pi^2-6)}\left[-24+\pi^2 \left(1-\frac{x^2}{\ell_0^2}\right)
\left( 3+\pi\frac{x}{\ell_0}\tan{\left(\frac{\pi x}{2 \ell_0}\right)}\right)\right],
\end{eqnarray}
we eventually obtain for $|x| < \ell_0$,
\begin{eqnarray}
n(x)&=& \frac{2}{\ell_0}
\sin^2{\left[\frac{\pi(x+\ell_0)}{2 \ell_0}\right]}\,
\exp{\left[\frac{1}{4(\pi^2-6)}\left[24+\pi^2 \left(\frac{x^2}{\ell_0^2}-1\right)
\left( 3+\pi\frac{x}{\ell_0}\tan{\left(\frac{\pi x}{2 \ell_0}\right)}\right)\right]\right]}\label{zetaOH}\\
\Gamma (\beta)&=&\frac{\sqrt{2 \pi} \lambda_D}{8 \ell_0}\exp{\left[-\beta E(\ell_0)\right]}.\label{GammaOH}
\end{eqnarray}
The normalized groundstate probability density,
$\Psi(x)=n(x)/\int \mathrm{d} z\, n(z)$ is compared  in Fig.~\ref{fig_pdf_OH}
with the exact Gaussian groundstate wavefunction.

\subsubsection{Anharmonic potentials}

\noindent Now, we consider more complicated potentials $V(z)=a_n z^n/n$,
where $n$ is an even integer. The associated Schr{\"o}dinger equation
depends only on the two parameters $\hbar^2/m$ and $a_n$. Only one
typical energy $\varepsilon_p=(\hbar^2/m)^{n/(n+2)} a_n^{2/(n+2)}$ and
one typical length $\ell_p=[\hbar^2/(m a_n) ]^{1/(n+2)}$ can be
built in terms of those parameters. Therefore, the
energy and the spatial extension of the ground state are
proportional to those typical energy and length respectively.
Again, formula~(\ref{geneconfin2}) can be
used to compute the ergodic density matrix elements in the low temperature
regime and for $|x| < \ell_0$. First, the function $E(\ell)$ reads
\begin{eqnarray} 
E(\ell)=\frac{\pi^2 \hbar^2}{8 m \ell^2}+\frac{a_n \ell^n}{n}R(n)  ,
\end{eqnarray}
where $R(n)=\int_{-1}^1 \mathrm{d}z \cos^2(\pi z/2) z^n$ depends on $n$. The
minimum of above expression, reached for
$\ell_0=\ell_p\left({\pi^2}/{4 R(n)}\right)^{1/(n+2)}$, is
\begin{eqnarray}  \label{EneAnha}
E(\ell_0)&=&a_n \ell_0^{\ n} R(n)\left[\frac{1}{2}+\frac{1}{n}\right].
\end{eqnarray}
We also find
\begin{eqnarray}
\varphi^{\pm}(\ell_0,x/\ell_0)&=&\frac{\pi}{2 n R(n)}\left[2\int_0^1\mathrm{d}z \, \sin(\pi z) z^{n+1}
-\int_{x/\ell_0}^1 \mathrm{d}z \, \sin(\pi z) z^n \right]\nonumber \\
&+& \frac{\pi}{2 n R(n)} \left(\frac{x}{\ell_0}\right) \tan\left(\frac{\pi x}{2 \ell_0}\right)\left[\int_0^1 \mathrm{d}z \, \cos^2\left(\frac{\pi
z}{2}\right) z^n-\frac{\ell_0}{x} \int_0^{x/\ell_0} \mathrm{d} z \, \cos^2\left(\frac{\pi
z}{2}\right)z^n\right]\label{pdfnOH}
\end{eqnarray}
which provides $n(x)$.

\bigskip

\begin{figure}[ht]
\psfrag{x}{\Large{$x/\ell_p$}} \psfrag{f}{\Large{$\Psi(x/\ell_p)$}}
\resizebox{0.4\textwidth}{!}{\includegraphics{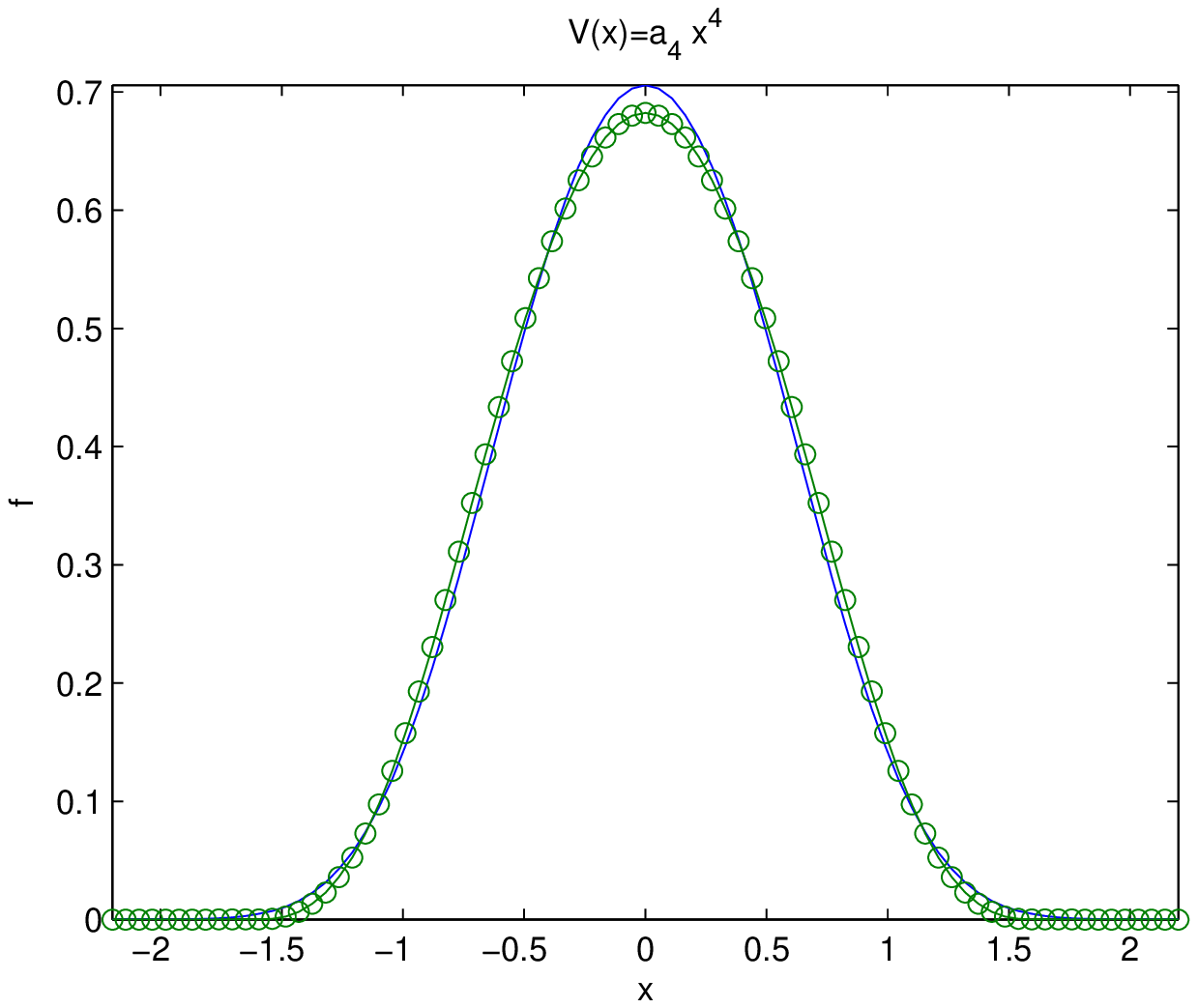}}
\resizebox{0.4\textwidth}{!}{\includegraphics{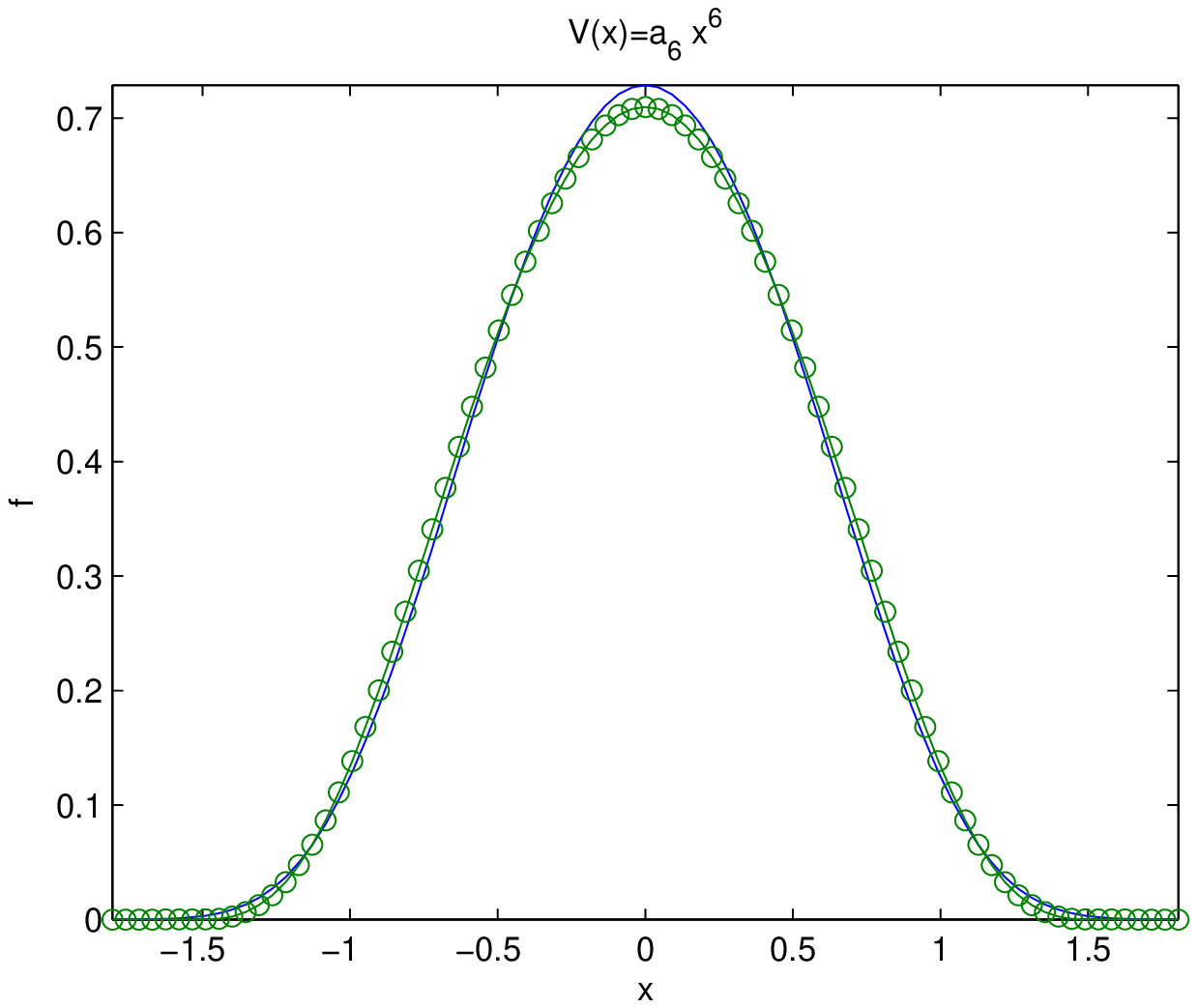}}
\resizebox{0.4\textwidth}{!}{\includegraphics{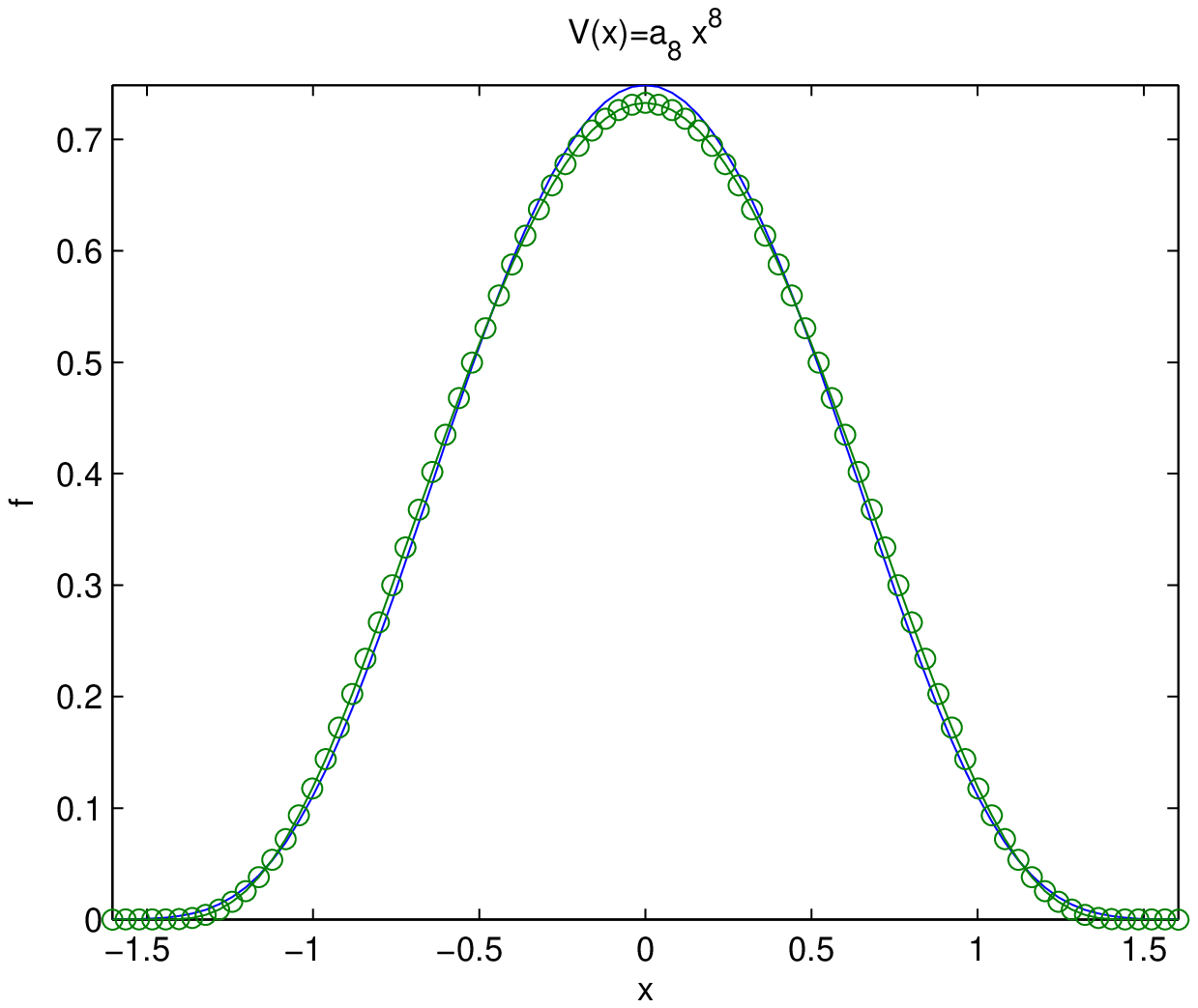}}
\resizebox{0.4\textwidth}{!}{\includegraphics{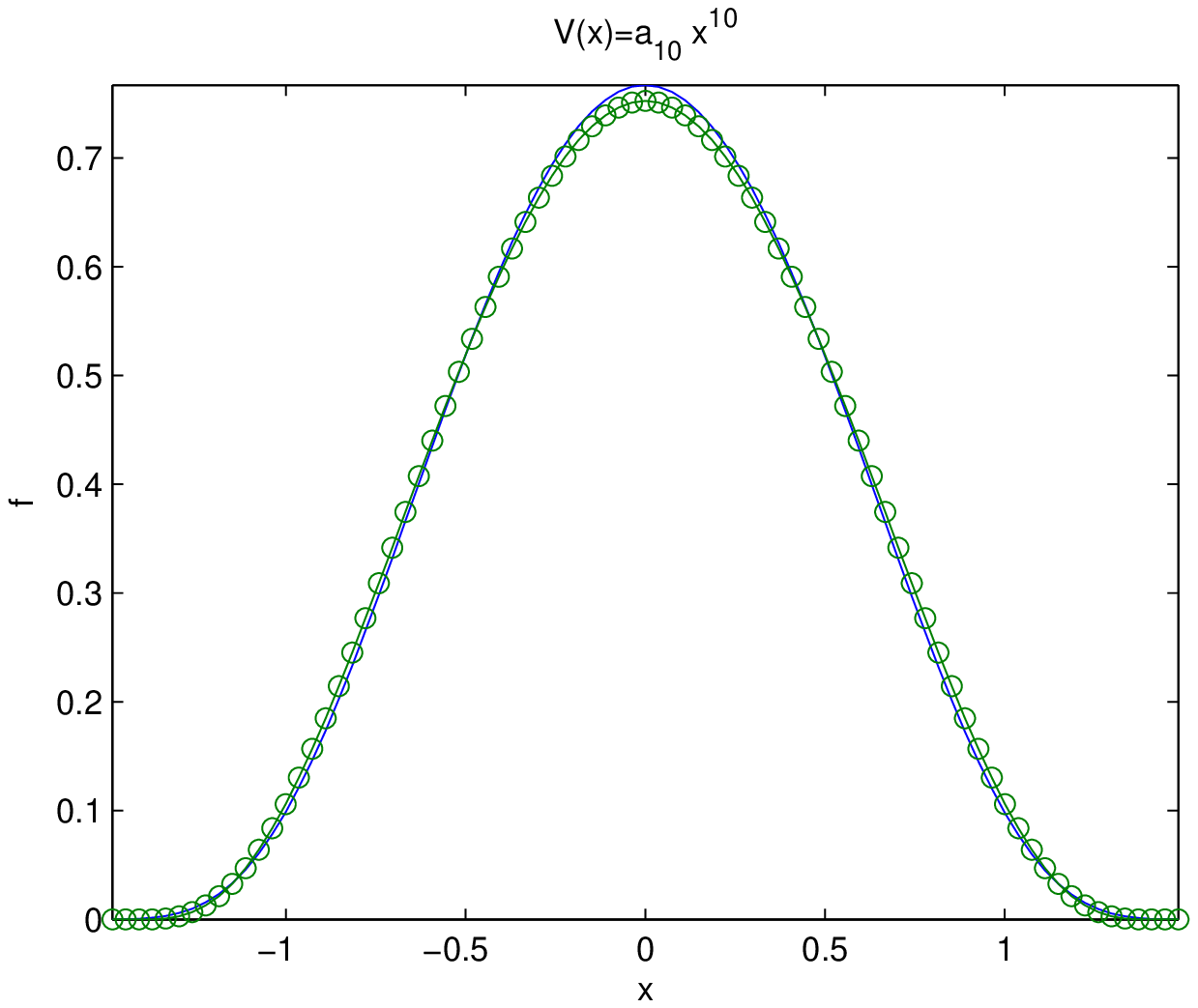}}
\caption{
  Probability density function of the ground state for anharmonic
  oscillators. The solid line is obtained through a numerical
resolution of Schr{\"o}dinger equation, while points correspond
to the ergodic approximation obtained from (\ref{pdfnOH}).}
\label{fig_pdfnOH}
\end{figure}

\noindent Two important comments are in order. The scaling properties of the
groundstate energy and extension are indeed recovered within
the ergodic approximation.
Table~\ref{tabletwo} presents a comparison with the numerical
resolution of Schr{\"o}dinger equation. The agreement, already
good for low values of the exponent $n$, becomes better when $n$ increases.
On the contrary, the semi-classical approximation completely
fails, since it provides a vanishing ground state energy (see Appendix
C). Comparisons for the groundstate pdf are presented in
Fig.~\ref{fig_pdfnOH}: the agreement is impressive.

\begin{table}
\begin{center}
\begin{tabular}{|c|c|c|c|c|c|}
\hline
${n}$     & $2$   & $4$   & $6$   &  $8$  &  $10$ \\
\hline \label{tab1}
{}      &               &               &               &               &               \\
$ \frac{\displaystyle E(\ell_0)-E_0}{\displaystyle E_0}$&        13\%    &       12\%    &       10\%    &       7.5\%   &       6\%     \\
\hline
\end{tabular}
\end{center}
\caption{Comparison of the ergodic groundstate energy $E(\ell_0)$ derived
by using formula~( \ref{EneAnha}), to the exact result $E_0$
(obtained by numerically solving Schr\"odinger equation). Integer
$n$ is the exponent of the anharmonic potential
}\label{tabletwo}
\end{table}

\subsection{High temperature regime}\label{app_ana2}

\noindent Either the Wigner-Kirkwood expansion or the semi-classical approximation become well-suited when
the temperature increases. Does the ergodic approximation
provides satisfactory results in the high temperature regime?
More precisely, what part of the Wigner-Kirkwood expansion
is accounted for within that approximation?
The corresponding calculations are carried out in Appendix D. We obtain
\begin{eqnarray}\label{ro1_ht}
\rho_{erg}(x,x,\beta) \!\!\!&\stackrel{\beta \to 0}{\sim}&\!\!\!\ \frac{e^{-\beta V(x)}}{\sqrt{2 \pi}\lambda_D}\left[ 1 -\beta\frac{\mathrm{d^2} V}{\mathrm{d}x^2}
\frac{\lambda_D^2}{12}+\left(\beta\frac{\mathrm{d} V}{\mathrm{d}x}\right)^2 \frac{\lambda_D^2}{24}A_0+o\left(\lambda_D^2\right) \right],
\end{eqnarray}
where $A_0$ is a numerical factor close to $0.8$. Thus,
the leading classical term is indeed recovered, while part of the
$\hbar^2$-correction is also correctly reproduced. At that
order $\hbar^2$, the discrepancy of (\ref{ro1_ht}) with
exact expansion (\ref{WK}), arises from correlations
between occupation times which are not taken into account
in the ergodic approximation. Nevertheless, that approximation
turns out to be also quite reasonable at high temperatures.

\subsection{Intermediate temperature regime}\label{app_ana3}

\noindent In this last section,  we present results obtained
trough the numerical implementation
of formula~(\ref{est2_1b}) at intermediate temperatures. As shown below, the ergodic approximation is particularly efficient in that
regime, or in other words, it provides a good interpolation between
the exact behaviours at high and low
temperatures. Here, we consider thermodynamics quantities, {\it i.e}
the internal energy $U=-\partial_{\beta}\left(\log\left(Z(\beta\right)\right)$
and the probability density function (PDF)
$\Psi(x,\beta)=\rho(x,x,\beta)/Z\left(\beta\right)$.
Comparisons are made with either exact results or other
familiar approximations.

\bigskip

\begin{figure}[ht]
\resizebox{0.47\textwidth}{!}{\includegraphics{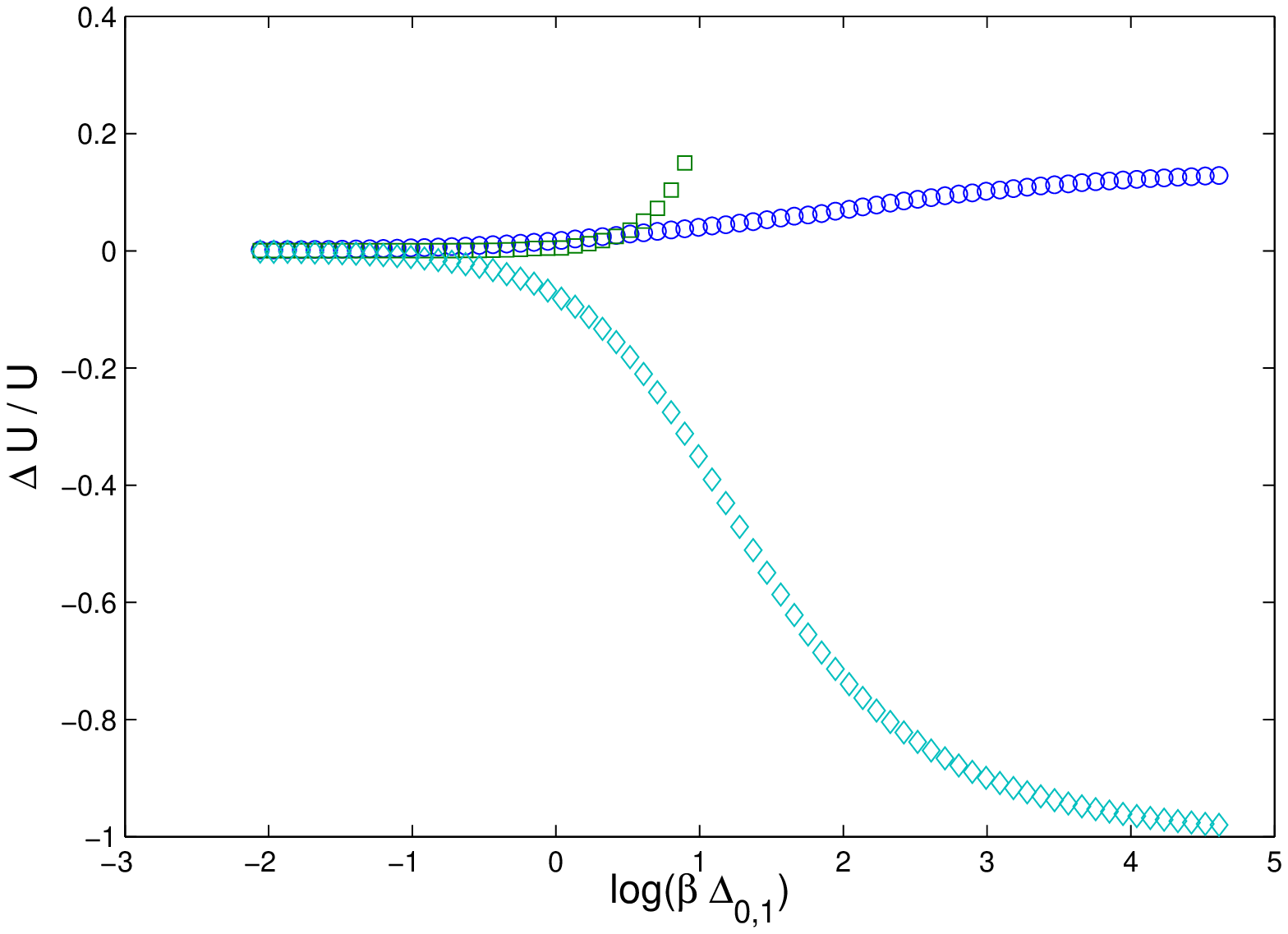}}
\resizebox{0.47\textwidth}{!}{\includegraphics{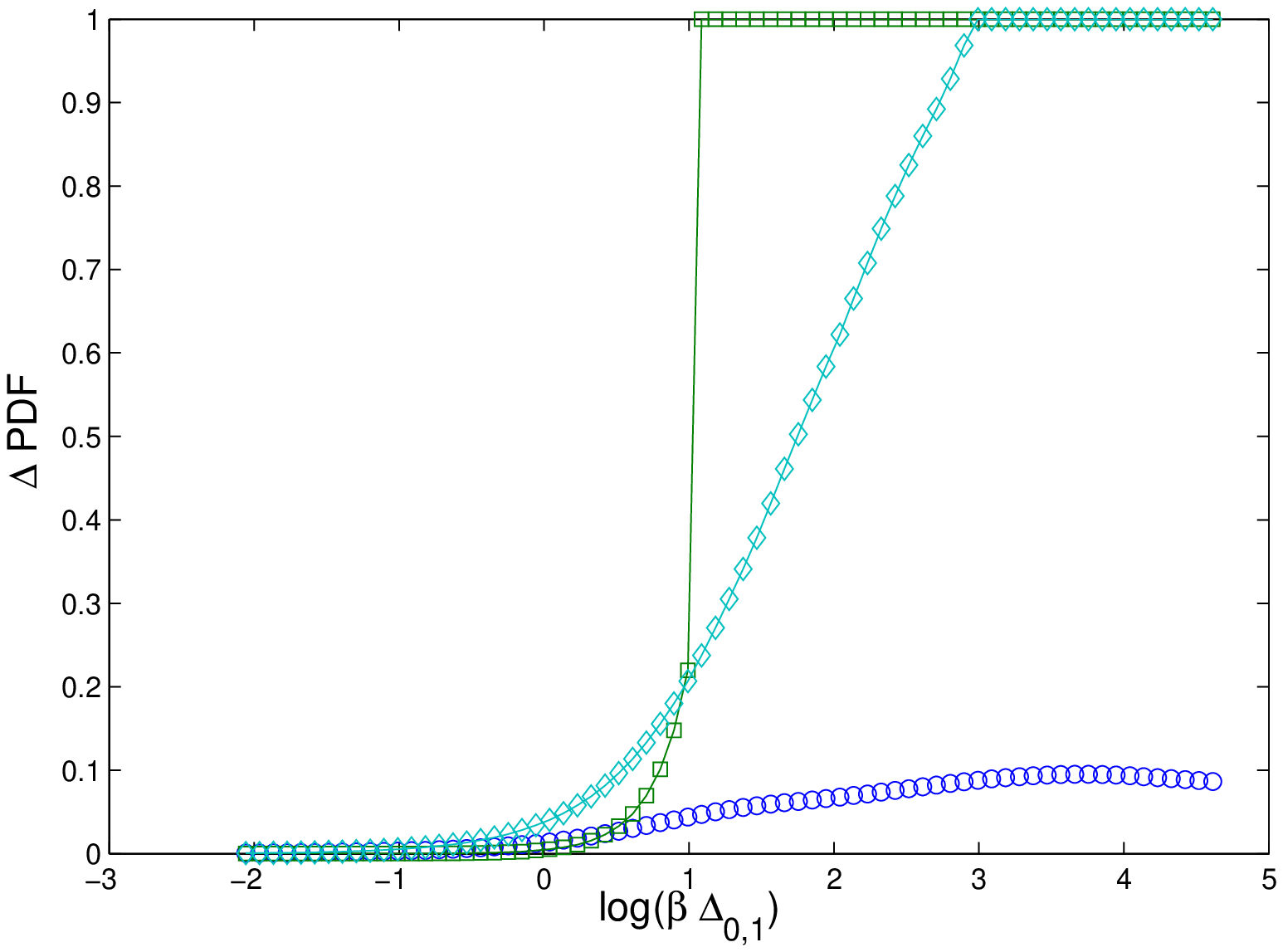}}
\caption{
  Relative error for the internal energy (left panel) and mean-square error
for the probability density function (right panel), as functions of
the temperature for the harmonic potential. 
Circles correspond to the ergodic approximation, square refers to Wigner-Kirkwood expansion truncated up to order
$\hbar^2$ and diamonds are the classical values. Quantity $\Delta_{0,1}$ is
  the energy gap between the ground state and the first excited one.}
\label{fig_compareOH}
\end{figure}

\begin{figure}[ht]
\resizebox{0.47\textwidth}{!}{\includegraphics{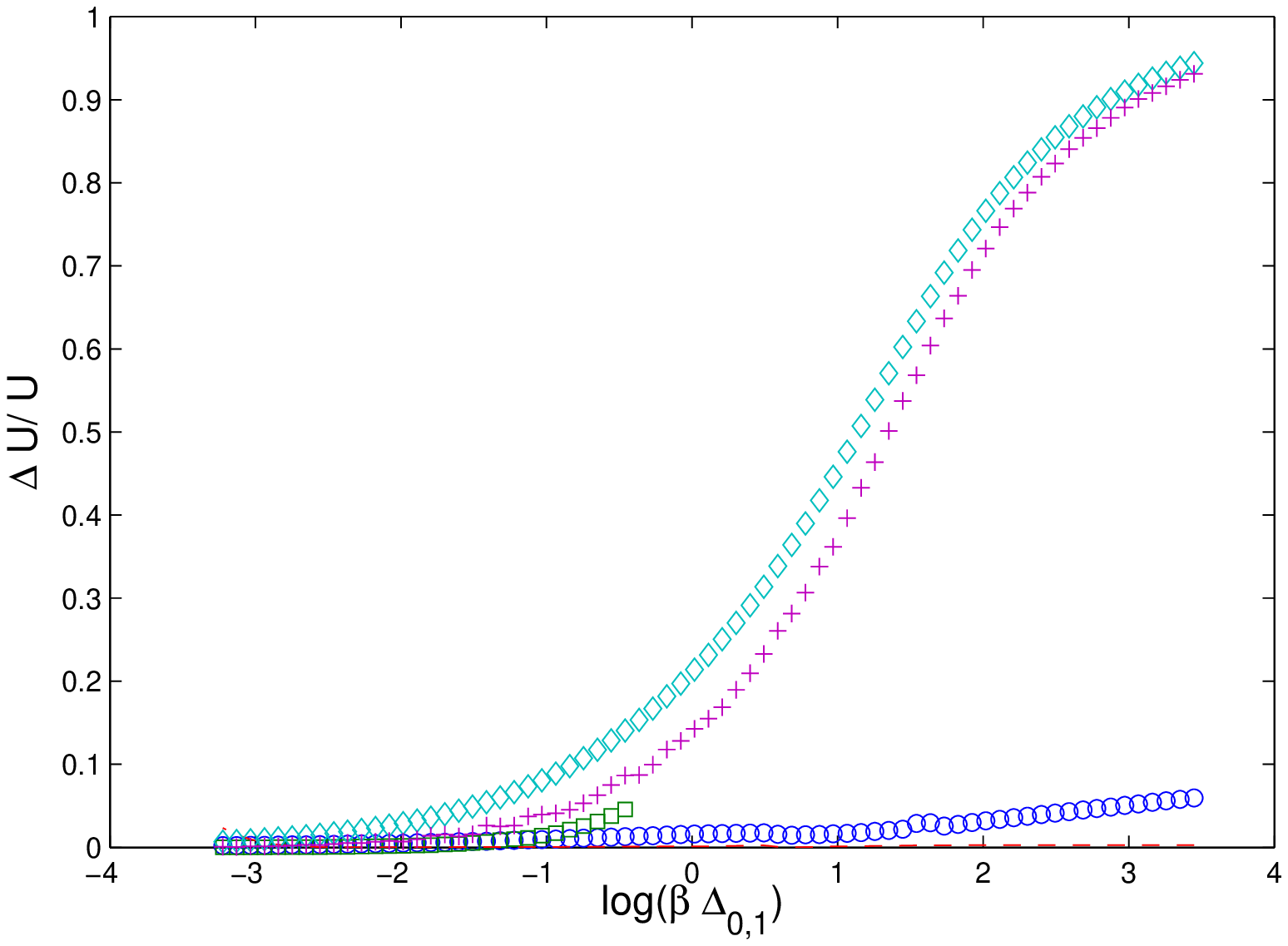}}
\resizebox{0.47\textwidth}{!}{\includegraphics{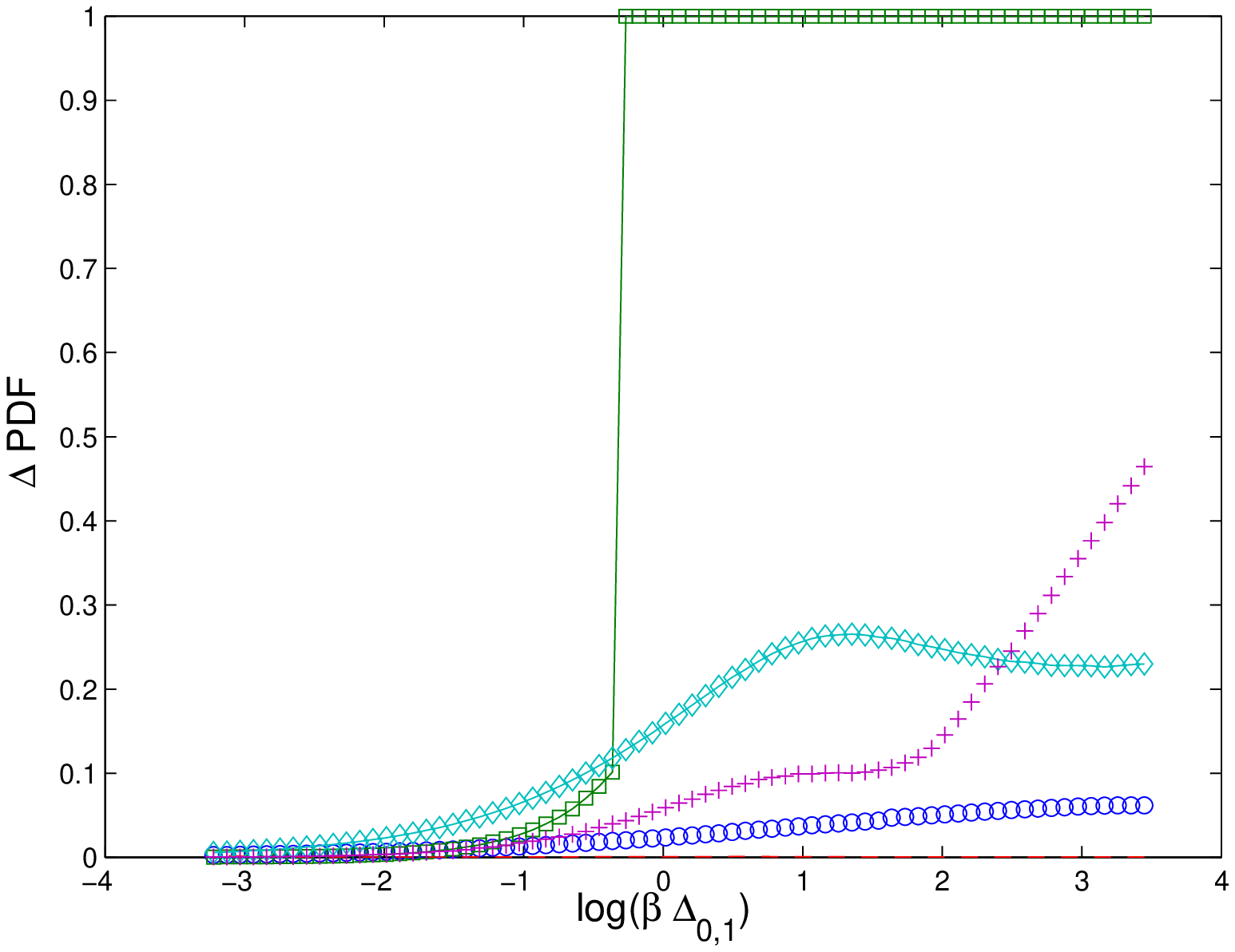}}
\caption{
  Relative error for internal energy (left panel) and mean-square error for the probability density function (right panel),
  as functions of the temperature for the potential $V(x)=x^{10}$. The exact results are obtained by using the spectral 
decomposition (\ref{scrho_dst_element}).
  Circles correspond to the ergodic approximation, square refers to Wigner-Kirkwood expansion truncated up to order $\hbar^2$,
crosses describe the semi-classical approximation and
  diamonds are the classical values. Quantity
 $\Delta_{0,1}$ is the energy gap between the ground state and the first excited one.}
\label{fig_compareX10}
\end{figure}

\noindent The ergodic approximation appears to be rather accurate for
estimating the considered thermodynamic quantities. For an
harmonic potential, it is well-known that the semi-classical approximation
turns out to be exact. As shown in figure \ref{fig_compareOH},
even in that case, the ergodic quantities are then close to the exact ones.
For anharmonic potentials, there always exists a large-temperature domain
where the ergodic approximation is significantly better than other
considered approximations. Figure \ref{fig_compareX10} shows results
for the potential $V(x)=x^{10}$. Above temperature domain
becomes larger when the anharmonicity of the potential increases.

\section*{Appendix A: Statistical weights of boxes}\label{generalexpressionforg}

\noindent Eigenstates and eigenvalues of the auxiliary Hamiltonian $H_{\ell}^0$
given in formula~(\ref{intermediatehamiltonian}) are obtained by solving the
Schr\"odinger equation
\begin{eqnarray}
\frac{\mathrm{d}^2 \phi^0_{\ell}}{\mathrm{d} z^2}=\frac{2 m V_0}{\hbar^2}
\left[1-\Theta \left(z+\ell\right)+\Theta
\left(z-\ell\right)-\frac{E}{V_0} \right]\phi^0_{\ell}(z).
\end{eqnarray}
The solutions are obtained by distinguishing three different regions. After introducing 
 dimensionless function $\psi(z)=\sqrt{\ell} \phi^0_{\ell}(z)$ we obtain
\begin{eqnarray}
\psi(x) &=&\left\{ \begin{array}{ll}
A \exp\left[K(z+\ell) \right], & \  \mbox{if}\ z \leqslant -\ell \\
B \sin\left[ k(z+\ell)+\delta \theta \right], & \  \ \forall z \in [-\ell, +\ell ] \ \ . \\
C \exp\left[-K(z-\ell) \right], & \  \mbox{if}\ z \geqslant \ell
  \end{array}
  \right.\label{expreeigenstate}
\end{eqnarray}
The continuity of the wave function and its first derivative in $\pm \ell$ leads to
the following four conditions
\begin{eqnarray}
A&=&B \sin\left( \delta \theta \right)
=\frac{k}{K}B \cos\left( \delta \theta \right),\\
C&=&B \sin\left[ 2 k \ell+\delta \theta \right]
=-\frac{k}{K}B \cos\left[ 2 k \ell +\delta \theta \right],
\end{eqnarray}
where we have introduced both wavevectors $k= (2 m E)^{1/2}/\hbar $
and $K=k({V_0}/E-1)^{1/2}$.
Rather than deriving the general expression for $A,C,k,\delta \theta$
as a function of $B$, it is sufficient to determine the asymptotic
expression for large values of potential barrier $V_0$. In the
infinite-$V_0$ limit, the dominant terms are
$\delta\theta\sim k/K$,
$A\sim Bk/K$,
$k\sim n \pi/(2 \ell)$ and
$C\sim(-1)^{(n-1)} B k/K$. Finally, the value $B=1$ is obtained
through the normalization condition.

\bigskip

\noindent Once those results are introduced in (\ref{expreeigenstate}), if we define the eigenvalues $E_n=\pi^2
\hbar^2 n^2/(8 m \ell^2)=E_1 n^2$ of the $n$-th eigenstates in the large-$V_0$-limit, we then obtain the
dimensionless eigenfunctions of the system in that asymptotic case
\begin{eqnarray}
\psi_n(z) &\stackrel{V_0\to\infty}{\sim}&\left\{ \begin{array}{ll}
\displaystyle \sqrt{\frac{E_1}{V_0}} n,  & \  \mbox{if}\ z=-\ell \\
\displaystyle \sin{\left[\frac{n \pi}{2 \ell} (z+\ell)\right]}, & \  \ \forall z \in ]-\ell, +\ell [   \ \ . \\
\displaystyle (-1)^{(n-1)} \sqrt{\frac{E_1}{V_0}} n, & \  \mbox{if}\ z = \ell.
  \end{array}
  \right . \label{arrayschro1}
\end{eqnarray}
By using the spectral decomposition (\ref{scrho_dst_element}), we obtain the density matrix elements of the free particle constrained to stay in the box $[-\ell,+\ell]$
\begin{eqnarray}
\rho_{\ell}^0\left(-\ell,x,\beta \right)=\rho_{\ell}^0\left(x,-\ell,\beta \right) &\stackrel{V_0\to\infty}{\sim}&\frac{1}{\ell}
\sqrt{\frac{E_1}{V_0}} \sum_{n=1}^{\infty} n \ \sin{\left[\frac{n \pi}{2 \ell} (z+\ell)\right]} \exp{\left[ -\beta E_1 n^2\right]},\\
\rho_{\ell}^0\left(\ell,x,\beta \right)=\rho_{\ell}^0\left(x, \ell,\beta \right) &\stackrel{V_0\to\infty}{\sim}&\frac{1}{\ell}
\sqrt{\frac{E_1}{V_0}} \sum_{n=1}^{\infty} (-1)^{n-1} n \sin{\left[\frac{n \pi}{2 \ell} (z+\ell)\right]}
\exp{\left[ -\beta E_1 n^2\right]},
\end{eqnarray}
which leads to the following expression for the statistical weight functions
\begin{eqnarray} \label{g_gene1}
g\pm (x,\ell,s,\beta)&=&\sqrt{2 \pi}  \frac{\pi^2 \lambda_D^3}{8 \ell^4}
S_g^{\pm}\left(\frac{x}{\ell},\frac{\pi\lambda_D \sqrt{s}}{2 \ell} \right)
S_g^{\pm}\left(\frac{x}{\ell},\frac{\pi\lambda_D \sqrt{1-s}}{2 \ell} \right),
\end{eqnarray}
where
\begin{eqnarray} \label{Sg_bt}
S_g^{\pm}(\alpha,y)=\sum_{n=1}^{\infty} n (\mp1)^{n-1}
\sin{\left[n \frac{\pi}{2}(\alpha+1)\right]}\exp{\left[-\frac{y^2}{2} n^2\right]}.
\end{eqnarray}
Those two functions satisfy the equality $g^+(x,\ell,s,\beta)=g^-(-x,\ell,s,\beta) \label{symg}$, the normalization condition
\begin{eqnarray}
\int_{|x|}^{\infty}\mathrm{d} \ell  \int_{0}^{1} \mathrm{d}s\left[g^-(x,\ell,s,\beta)+g^+(x,\ell,s,\beta)\right] =1,
\end{eqnarray}
and they take a simple  asymptotic form in the small-temperature regime
\begin{eqnarray}\label{gbt}
g\pm (x,\ell,s,\beta)&\stackrel{\lambda_D\gg\ell}{\sim}& \sqrt{2 \pi}  \frac{\pi^2 \lambda_D^3}
{8 \ell^4} \sin^2{\left[\frac{\pi(x+\ell)}{2 \ell}\right]}
\exp{\left[-\frac{\pi^2 \lambda_D^2}{8 \ell^2}\right]}.
\end{eqnarray}
The asymptotic expression of $g\pm$ in the high-temperature regime is obtained by taking the Poisson transform
of expression (\ref{Sg_bt}). According to Poisson formula, if the Fourier transform of a function $f$
exists, then, for all values of $\Delta y$
\begin{eqnarray}\label{poisson}
\Delta y \sum_{n= -\infty}^{\infty} f(n \Delta y) = \sum_{m=-\infty}^{+\infty} \int_{-\infty}^{+\infty} \mathrm{d} z \, f(z)
\ e^{-\mathrm{i}m 2\pi z / \Delta y}.
\end{eqnarray}
Application of that transformation to expression (\ref{Sg_bt}) provides
\begin{eqnarray}\label{Sg_ht}
S_g^{\pm}(\alpha,y)=\left( \frac{\pi}{2 y^2} \right)^{3/2} \sum_{m=-\infty}^{+\infty}
(1\mp\alpha-4m) \exp\left[-\frac{\pi^2}{8 y^2}(1\mp\alpha-4m)^2\right].
\end{eqnarray}
Therefore, in the high temperature regime, $g\pm$ reads
\begin{eqnarray}\label{ght}
g^-(x,\ell,s,\beta)=g^+(-x,\ell,s,\beta)&\stackrel{\lambda_D\ll\ell}{\sim}&
\sqrt{\frac{2}{ \pi}}  \frac{\ell^2}{\lambda_D^3}
\frac{1}{\sqrt{s(1-s)}}\frac{\left(1+x/\ell\right)^2}{s(1-s)} \exp\left[-\frac{\ell^2}{2 \lambda_D^2 } \frac{(1+ x/\ell)^2}{s(1-s)}\right].
\end{eqnarray}

\section*{Appendix B:  Derivation of the mean occupation time by
using the operator method}

\noindent Expression~(\ref{Phigene}) for the averaged occupation
time $\left\langle\theta(x')\right\rangle_{\omega} $ can be alternatively
computed by introducing the effective Hamiltonian
$H_{x'}=H_{\ell}^0+V_{x'}$,
with $\langle x | V_{x'}= \varepsilon \delta(x-x') \langle x
|$.  Indeed, using a Taylor expansion of the exponential, the density
matrix elements associated to that hamiltonian can be first written as
\begin{eqnarray}
\frac{\langle x_i |\exp{\left[-\beta H_{x'}\right]}| x_f \rangle}
{\langle x_i |\exp{\left[-\beta H_{\ell}^0 \right]}| x_f \rangle}
&=& \frac{\langle x_i |
\exp{\left[-\beta H_{\ell}^0\right]}| x_f \rangle
+\varepsilon \langle x_i | \partial_{\varepsilon} \left[
\exp{\left[-\beta H_{x'}\right]}
\right]_{\varepsilon = 0} | x_f \rangle +O\left(\varepsilon^2\right) }{\langle x_i |\exp{\left[-\beta
H_{\ell}^0 \right]}| x_f \rangle}  \\
&=& 1+ \displaystyle
\varepsilon \frac{\langle x_i | \partial_{\varepsilon} \left[
\exp{\left[-\beta H_{x'}\right]} \right]_{\varepsilon = 0} |
x_f \rangle}{\langle x_i |\exp{\left[-\beta
H_{\ell}^0 \right]}| x_f \rangle}
+O\left(\varepsilon^2\right).
\end{eqnarray}
The Dyson derivation formula~(\ref{formuledyson}) and the
introduction of a closure relation lead to
\begin{eqnarray}
\frac{\langle x_i |\exp{\left[-\beta H_{x'}\right]}| x_f \rangle}
{\langle x_i |\exp{\left[-\beta H_{\ell}^0 \right]}| x_f \rangle}
&=& 1- \displaystyle
\beta \varepsilon \frac{\displaystyle \int_0^1 \mathrm{d} s
\int \mathrm{d} z \ \langle x_i |\exp{\left[-\beta(1-s)
H_{\ell}^0 \right]}| z \rangle
\delta(z-x') \langle z |\exp{\left[-\beta s
H_{\ell}^0 \right]}| x_f \rangle
}{\displaystyle \langle x_i |\exp{\left[-\beta
H_{\ell}^0 \right]}| x_f \rangle}
+O\left(\varepsilon^2\right) \\
&=& 1- \displaystyle
 \beta\varepsilon \frac{ \displaystyle \int_0^1 \mathrm{d} s \
\langle x_i |\exp{\left[-\beta(1-s) H_{\ell}^0 \right]}| x' \rangle
\langle x' |\exp{\left[-\beta s H_{\ell}^0 \right]}| x_f \rangle
}{ \displaystyle \langle x_i |\exp{\left[-\beta
H_{\ell}^0 \right]}| x_f \rangle} +O\left(\varepsilon^2\right).
\end{eqnarray}
Above relation can be further simplified by introducing the wave
functions $\phi^0_n(x)=\ell^{-1/2} \psi_n(x)$, given in Eq.~(\ref{arrayschro1}), of the
constrained Hamiltonian $H_{\ell}^0$, and its
associated eigenvalues $E_n=E_1 n^2$.  We obtain
\begin{eqnarray}
\frac{\langle x_i |\exp{\left[-\beta H_{x'}\right]}| x_f \rangle}
{\langle x_i |\exp{\left[-\beta H_{\ell}^0 \right]}| x_f \rangle}
\!\!\!\!&=& 1-
\frac{\beta\varepsilon }{\ell}\frac
{ \displaystyle \int_0^1 \mathrm{d} s
\left(\sum_{k}\psi_k(x_i) \psi_k^{*}(x')e^{-\beta(1-s) E_k} \right)
\left(\sum_{n}\psi_n(x')\psi_n^{*}(x_f) e^{-\beta s    E_n} \right)}
{\displaystyle \sum_{n}\psi_n(x_i) \psi_n^{*}(x_f) e^{-\beta E_n}} +O\left(\varepsilon^2\right)\\
&=&\!\!\!1-\!\!
\frac{\beta\varepsilon }{\ell}  \frac{ \displaystyle \sum_{n} \psi_n(x_i) |
\psi_n(x') |^2 \psi_n^{*} (x_f)
e^{-\beta E_n }\!\!+ \!\! \sum_{n \neq k}
\psi_k(x_i) \psi_k^{*}(x') \psi_n(x') \psi_n^{*} (x_f)
\frac{e^{-\beta E_k}-e^{-\beta E_n }}{\beta(E_n-E_k)} }{\displaystyle
\sum_{n} \psi_n(x_i) \psi_n^{*}(x_f)\exp{\left[-\beta E_n\right]}}\!\!+\!O\left(\varepsilon^2\right).\label{resulviaoperator}
\end{eqnarray}
The left-hand-side can be rewritten by using the Feynman-Kac formula as
\begin{eqnarray}
\frac{\langle x_i |\exp{\left[-\beta H_{x'}\right]}| x_f \rangle}
{\langle x_i |\exp{\left[-\beta H_{\ell}^0 \right]}| x_f \rangle}
&=&\frac{ \displaystyle\int_{{\omega}}{ \mathcal{D}_W (\xi)
\exp{\left[-\beta \varepsilon\int_0^1 \mathrm{d} s \ \delta (x'- x_i(1-s)-s x_f - \lambda_d \xi(s))\right] }}}
{\displaystyle\int_{{\omega}}\,\mathcal{D}_W(\xi)}.
\end{eqnarray}
(The effect of part $H_{\ell}^0$ of the Hamiltonian
is to restrict the integration domain to the subset $\omega$.) 
Finally, using a Taylor expansion of the exponential leads to
\begin{eqnarray}
\frac{\langle x_i |\exp{\left[-\beta H_{x'}\right]}| x_f \rangle}
{\langle x_i |\exp{\left[-\beta H_{\ell}^0 \right]}| x_f \rangle}
&=&\frac{\displaystyle \int_{{\omega}}{ \mathcal{D}_W (\xi)
\left(1-\beta \varepsilon\int_0^1\mathrm{d} s \
\delta (x'- x_i(1-s)-s x_f  - \lambda_d \xi(s)) +O\left(\varepsilon^2\right) \right)
}}{\displaystyle\int_{{\omega}}\,\mathcal{D}_W(\xi)}\\
&=&1-\beta \varepsilon \frac{\displaystyle \int_{{\omega}}
{ \mathcal{D}_W (\xi) \, \int_0^1 \mathrm{d} s \ \delta (x'- x_i(1-s)-s x_f  - \lambda_d \xi(s)) }}
{\displaystyle\int_{{\omega}}\,\mathcal{D}_W(\xi)} +O\left(\varepsilon^2\right) \\
&=&1-\beta \varepsilon \, \left\langle
\theta(x')\right\rangle_{\omega} +O\left(\varepsilon^2\right).\label{resulviaFeynmanKAc}
\end{eqnarray}
Combining Eq.~(\ref{resulviaoperator}) and~(\ref{resulviaFeynmanKAc}) leads to the final result~(\ref{Phigene}).

\section*{Appendix C:  Semi-classical approximation for potentials
$V(z)=a_n z^n$ in the low-temperature regime}

\noindent In this appendix, we present the semi-classical analysis
for potentials $V(z)=a_n z^n$, where $n$ is an even
integer and $a_n$ a positive real constant. The semi-classical density matrix elements are given by Dashen Formula~\cite{dashenformula}
\begin{eqnarray}\label{dashgene}
\rho_{sc}(x,x,\beta) = \frac{1}{\sqrt{2 \pi \hbar}} \exp{\left[-\frac{S_c[.]}{\hbar}\right]} \left| \frac{\partial P_0(x,x_f)}{\partial x_f}\right|_{(x_f=x)}^{1/2},
\end{eqnarray}
where $S_c[.]$ is the classical Euclidian action of the trajectory
starting from the initial point $x$ and coming back to that point in
a time $\beta \hbar$, and $P_0(x_i,x_f)$ is its initial momenta.

\bigskip

\noindent The classical Euclidian action for the particle moving in potential $-V$,
\begin{eqnarray}
S_c[z]=\int_{0}^{\beta\hbar} \mathrm{d} t \ \left[\frac{m}{2} \dot z ^2+V(z(t)) \right], \label{Sc1}
\end{eqnarray}
can be rewritten, if one introduces the classical energy $ E=(m\dot
z^2)/2-V(z)$ and uses that the exponent $n$ is even,
as
\begin{eqnarray}
S_c[z]
=\beta \hbar E +2 \int_{0}^{\beta\hbar} \!\!\! \mathrm{d} t  \, V(z(t)). \label{Sc1suite}
\end{eqnarray}
For negative energy values $E$, we introduce  characteristic lengths $z_m=(-E/a_n)^{1/n}$ and $\ell_p=[\hbar^2/(m a_n) ]^{1/(n+2)}$ 
and characteristic energy $\varepsilon_p= an \, \ell_p ^n$. This leads to
\begin{eqnarray}\label{zpoint}
 |\dot z|
=\left(\frac{2 \varepsilon_p}{m}\right)^{1/2} \left(\frac{z_m}{\ell_p}\right)^{n/2} \left[\left(\frac{z}{z_m}\right)^n-1\right]^{1/2}. \label{z1}
\end{eqnarray}
 Expression  (\ref{Sc1suite}) can thus be rewritten as
\begin{eqnarray}
S_c[z]&=&\beta \hbar \varepsilon_p\left(\frac{z_m}{\ell_p}\right)^n \left[-1 +2 \sqrt{\frac{2}{\beta \varepsilon_p}}
\left| \frac{z_m}{\lambda_D} \right|  \left(\frac{\ell_p}{z_m}\right)^{n/2} \!\!\! \left(\frac{x}{z_m}\right)^{(n+2)/2} \!\!\! F_n(x/z_m) \right] \label{Sutil},
\end{eqnarray}
with $F_n(y)=y^{-(n+2)/2)} \int_0^y \mathrm{d} u \, u^n/\sqrt{u^n-1}$.
The turning point $z_m$ is determined through the constraint that the
duration of the motion from $z_m$ to $x$ is precisely $\beta \hbar
/2$.

\bigskip

\noindent In the more general case, if $x_i=z(0)$ is the starting point of the trajectory and $x_f=z(\beta\hbar)$ is it's ending point, the constrain
over time is
\begin{eqnarray}
\beta \hbar &=& \int_{x_i}^{z_m} \left| \frac{\mathrm{d} z}{\dot z} \right|+\int_{z_m}^{x_f} \left| \frac{\mathrm{d} z}{\dot z} \right|.
\end{eqnarray}
Using expression (\ref{zpoint}), that constrain can be rewritten as
\begin{eqnarray}\label{contrainte}
\left(2 \beta \varepsilon_p \right)^{1/2} =\left| \frac{z_m}{\lambda_D}\right| \left(\frac{\ell_p}{z_m}\right)^{n/2}\left[ I_n(x_i/z_m) +
I_n(x_f/z_m) \right],
\end{eqnarray}
with $I_n(y)=\int_0^y \mathrm{d} u \,/\sqrt{u^n-1}$.
In the low temperature regime (i.e. when $\beta \hbar$ tends to
infinity), we can derive the asymptotic expression of the
characteristic length $z_m$. Setting $z_0=z_m(x,x)$ the solution of implicit equation (\ref{contrainte}) for $x_i=x_f=x$, we 
calculate $z_m(x,x_f)=z_0+\delta z$ if $x_f=x+\delta x$ is very close to $x$. The linearization of
equation (\ref{contrainte}) then leads to
\begin{eqnarray}\label{deltazsurz}
\frac{\delta z}{z_0}=\frac{\delta x}{x} \frac{1}{\displaystyle 2+(n-2)\left(\frac{z_0}{x}\right) I_n(x/z_0) \left[(x/z_0)^n-1\right]^{1/2}}.
\end{eqnarray}

\bigskip

\noindent The quadratic case  $n=2$ is singled out since
equation (\ref{contrainte}) can be analytically solve for all values of $\beta$. We obtain
\begin{eqnarray}
z_0=\frac{x}{\displaystyle \cosh\left[\frac{\lambda_D}{\ell_p}(\beta \varepsilon_p/2)^{1/2}\right]}\quad\mbox{and}\qquad
S_c[.]=m x^2 \sqrt{\frac{2 a_2}{m}}\tanh\left( \frac{\beta \hbar}{2}\sqrt{\frac{2 a_2}{m}} \right)\label{S2}
\end{eqnarray}
which ultimately provides the exact density matrix.
The non-quadratic cases ($n>2$)  are more interesting. At low
temperatures, we find
\begin{eqnarray}
 z_m \sim \ell_p \left[ \frac{2 I_n^2 \ell_p^2}{\beta \varepsilon_p \lambda_d^2}\right]^{1/(n-2)}\quad\mbox{and}\qquad
S_c[.]\sim 2 |x| \sqrt{ 2 m \varepsilon_p}  \left(\frac{|x|}{\ell_p}\right)^{n/2} F_n , \label{Sn}
\end{eqnarray}
where $I_n=\lim_{y\to  \infty}I_n(y)$ and $F_n=\lim_{y\to  \infty} F_n(y)$ are numbers of order $1$, estimated numerically (see table~\ref{tableappendixB}).
\begin{table}[!ht]
\begin{center}
\begin{tabular}{|c|c|c|c|c|}
\hline
n       &  $4$          & $6$           &  $8$          &       $10$    \\
\hline
$F_n$   &       0.33    &       0.25    &       0.20    &       0.16    \\
\hline
$I_n$   &       1.31    &       0.7     &       0.48    &       0.37    \\
\hline
\end{tabular}
\end{center}
\caption{Value of $F_n$ and $I_n$ for different even exponents $n$ of the potential $V(z)=a_nz^n$. }\label{tableappendixB}
\end{table}

\bigskip

\noindent To achieve the calculation of formula (\ref{dashgene}), we have to compute the  partial derivative of the initial momenta with
respect to the ending point of the classical path. Energy
conservation reads
\begin{eqnarray}
P_0^2(x,x_f)=\left(2 m \varepsilon_p\right) \left(\frac{z_m}{\ell_p}\right)^n \left[ \left(\frac{x}{z_m}\right)^n-1 \right].
\end{eqnarray}
For $z_m$ close to $z_0$, we linearize $P_0(x,x_f)=P_0(x,x)+\delta P_0(x,x_f)$, with the result
\begin{eqnarray}
\delta P_0(x,x_f)=\left(\frac{2 m \varepsilon_n}{ (x/z_0)^n -1}\right)^{1/2} \frac{n \, \delta z}{2 z_0},
\end{eqnarray}
so
\begin{eqnarray}
\left|\frac{\partial P_0(x,x_f)}{\partial x_f}\right|_{(x_f=x)}=\frac{ (n/2|x|) \sqrt{2 m \varepsilon_p} (z_0/\ell_p)^{n/2}}
{\displaystyle 2 \left[
(x/z_0)^n-1\right]^{1/2} +(n-2) \frac{z_0}{x} I_n(x/z_0) \left[(x/z_0)^n-1\right] }.
\end{eqnarray}
In the quadratic case, we find
\begin{eqnarray}\label{partial_P_2}
\left|\frac{\partial P_0(x,x_f)}{\partial x_f}\right|_{(x_f=x)}\!\!\!\!=\frac{\sqrt{2 m a_2}}{\sinh\left[ (\beta \hbar/m) \sqrt{2 m
a_2} \right]}.
\end{eqnarray}
For non-quadratic cases ($n>2$), we have the low-temperature behaviours
\begin{eqnarray}\label{partial_P_n}
\left|\frac{\partial P_0(x,x_f)}{\partial x_f}\right|_{(x_f=x)}\!\!\!\! \sim \frac{2 n}{(n-2)} \left(\frac{\ell_p}{x}\right)^n \frac{ I_n \hbar \,
\ell_p
}{(2 \beta \varepsilon_p)^{1/2} \lambda_D^3}\left[ \frac{2 I_n^2 \ell_p^2}{\beta \varepsilon_p \lambda_d^2}\right]^{(n+2)/(2n-4)}.
\end{eqnarray}
Finally, by combining (\ref{dashgene}), (\ref{partial_P_2}) and (\ref{S2}), we obtain the exact density matrix for the harmonic potential
\begin{eqnarray}
\rho_{sc}(x,x,\beta)=\frac{(2 m a_2)^{1/4}}{(2 \pi \hbar)^{1/2}}\frac{\exp\left[-m x^2 \sqrt{\frac{2 a_2}{m \hbar^2}}\tanh\left(
\frac{\beta \hbar}{2 m}\sqrt{2 m a_2} \right) \right]}{\sinh\left[ (\beta \hbar/m) \sqrt{2 m
a_2} \right]}.
\end{eqnarray}
For anharmonic potentials, the
semi-classical density matrix behaves as
\begin{eqnarray}
\rho_{sc}(x,x,\beta)\sim \frac{1}{\lambda_D} \sqrt{\frac{n}{\pi (n-2)}} \left(\frac{\ell_p}{|x|}\right)^{n/2}\!\!\! \left(\frac{ I_n \,
\ell_p}{(2 \beta \varepsilon_p)^{1/2} \lambda_D} \right)^{1/2}
\left[ \frac{2 I_n^2 \ell_p^2}{\beta \varepsilon_p \lambda_d^2}\right]^{(n+2)/(4n-8)}\!\!\!\!\!\!\!\!\!
\exp\left[-2 |x| \sqrt{ 2 m \varepsilon_p/ \hbar^2}  \left(\frac{|x|}{\ell_p}\right)^{n/2} \!\!\!\! F_n \right].
\end{eqnarray}
When $\beta \to 0$ that expression does not reproduce the expected exponential decay with respect to temperature.

\section*{Appendix D: Ergodic density matrix at high temperatures.}

\noindent Let us focus on the calculation of $\rho_{erg}(x,x,\beta)$ 
in the high-temperature regime. We obtain it's the asymptotical
form for values of $x$ such that
$|x|/\lambda_D\gg 1$. For simplification purposes, we restrict our
analysis to negative values of $x$. In that case, using 
asymptotic expression (\ref{ght}), we realize that we can omit the
paths taken into account by the density of measure
$g^{+}(x,\ell,s,\beta)$ since their contributions are negligible (on the
contrary, for positive values of $x$, we can omit the contribution of
paths implied in $g^{-}(x,\ell,s,\beta)$).Thus, expression
(\ref{est2_1b}) reduces to
\begin{eqnarray}\label{forme1}
\rho_{erg}(x,x,\beta) \stackrel{\beta \to 0}{\sim}
\int_{-x}^{+\infty} \!\!\!\!\!\!\!\! &\mathrm{d} \ell&
\int_{0}^1 \!\! \mathrm{d} s\,
\frac{\left(\ell+x\right)^2}{\pi \lambda_D^4 s^{3/2}(1-s)^{3/2}} \exp\left[-\frac{(\ell+ x)^2}{2 \lambda_D^2 s(1-s)}\right] \nonumber \\
&\times&\exp{\left(-\beta \int_{-1}^{1} \!\!\!\!\mathrm{d} \alpha'
\left[s \Phi \left(\frac{x}{\ell},\alpha',\frac{\pi \lambda_D \sqrt{s}}{2 \ell}\right)+
[1-s] \Phi \left(\frac{x}{\ell},\alpha',\frac{\pi \lambda_D \sqrt{1-s}}{2 \ell}\right)
\right] V\left( \alpha' \ell \right)\right)}.
\end{eqnarray}
That expression clearly shows that paths with extension larger
than $\lambda_D$ with respect to $x$ give a vanishing
contribution to the integrals. Since we are interested in the limit
$|x|/\lambda_D\gg 1$, we see that for leading paths
$x/\ell=-1+\Delta \alpha$ with $\Delta \alpha\ll 1$. By starting from
 (\ref{Phiht1})-(\ref{Phiht2}), we find the asymptotic form of the function
$\Phi(\alpha,\alpha',y)$ in that parameter range,
\begin{eqnarray}
\Phi(\-1+\Delta \alpha,-1+\Delta \alpha',y) &\stackrel{\Delta \alpha\ll1}{\sim}&\left\{ \begin{array}{ll}
\displaystyle \Phi_1\left(\Delta \alpha,\Delta \alpha',y\right)=
\frac{1}{\Delta \alpha}\left[ 1-\exp\left( -\frac{\pi^2}{2 y^2}
(\Delta \alpha'+\Delta \alpha) \Delta \alpha'\right) \right]   &, \ \forall \Delta
\alpha'< \Delta\alpha\\ \\
\displaystyle \Phi_2(\Delta \alpha,\Delta \alpha',y)=\frac{2}{\Delta
\alpha}\exp\left( -\frac{\pi^2}{2 y^2} \Delta\alpha'^2 \right)
\sinh\left(\frac{\pi^2}{2 y^2}\Delta\alpha \Delta\alpha'\right)   &, \ \forall \Delta
\alpha'> \Delta\alpha
  \end{array}
  \right . \label{arraysphi}
\end{eqnarray}
Then, the integral appearing in the exponential of (\ref{forme1})
can be calculated by using a Taylor expansion of the potential around the position $x$. We obtain
\begin{eqnarray}
\int_{-1}^{1} \!\!\!\!\mathrm{d} \alpha'
\left[s \Phi \left(\frac{x}{\ell},\alpha',\frac{\pi\lambda_D\sqrt{s}}{2\ell}\right)+
[1-s] \Phi \left(\frac{x}{\ell},\alpha',\frac{\pi\lambda_D\sqrt{1-s}}{2\ell}\right)
\right] V\left( \alpha' \ell \right)=V(x) + \sum_{k=1}^{\infty}\left[V_k(\ell,s)+V_k(\ell,1-s)\right]
\end{eqnarray}
with
\begin{eqnarray}
V_k(\ell,u)=u \frac{\ell^k}{k ! } \frac{\mathrm{d}^k V}{{\mathrm{d}x}^k}\int_0^{2}
\mathrm{d}z \Phi\left(\-1+\Delta \alpha,-1+z,\frac{\pi\lambda_D\sqrt{u}}{2\ell}\right)(z-\Delta \alpha)^k.
\end{eqnarray}
The explicit forms of the first terms are
\begin{eqnarray}
V_1(\ell,u)&=&-\frac{1}{2}\frac{\mathrm{d} V}{\mathrm{d}x}\left[ u
(\ell+x)- u^{3/2} \sqrt{\frac{\pi}{2}}\lambda_D
\exp\left(\frac{(\ell+x)^2}{2 \lambda_D^2
u}\right)\text{erfc}\left(\frac{\ell+x}{\lambda_D \sqrt{2 u}}
\right) \right], \\ \nonumber \text{and} \ \ \
V_2(\ell,u)&=&\frac{1}{2}\frac{\mathrm{d}^2
V}{{\mathrm{d}x}^2}\left[ \frac{1}{3}u (\ell+x)^2+\frac{1}{2}u^2
\lambda_D^2- u^{3/2} \sqrt{\frac{\pi}{2}} \lambda_D(\ell+x)
\exp\left(\frac{(\ell+x)^2}{2 \lambda_D^2 u}\right)\text{erfc}\left(
\frac{\ell+x}{\lambda_D \sqrt{2 u}} \right) \right],
\end{eqnarray}
where $\text{erfc}(z)$ is the complementary error function.
Expression (\ref{forme1}) can be formally rewritten as
\begin{eqnarray}\label{forme2}
\rho_{erg}(x,x,\beta) \!\!\!&\stackrel{\beta \to 0}{\sim}&\!\!\!
\int_{-x}^{+\infty} \!\!\!\!\!\!\!\! \mathrm{d} \ell
\int_{0}^1 \!\!
\frac{\mathrm{d} s\, \left(\ell+x\right)^2}{\pi \lambda_D^4 s^{3/2}(1-s)^{3/2}}
\exp\left(-\frac{(\ell+ x)^2}{2 \lambda_D^2 s(1-s)}-\beta V(x)-\beta \sum_{k=1}^{\infty}\left[V_k(\ell,s)+V_k(\ell,1-s)\right] \right).
\end{eqnarray}
Since the temperature is high, part of the exponential implying derivatives of the potential can be expanded in power series.
The truncation of these series up to second order with respect to $\lambda_D$ provides
\begin{eqnarray}\label{forme3}
\rho_{erg}(x,x,\beta) =
\frac{e^{-\beta V(x)}}{\sqrt{2 \pi} \lambda_D} \int_{-x}^{+\infty} \!\!\!\!\!\!\!\! &\mathrm{d}\ell&
\int_{0}^1 \!\! \mathrm{d} s\,\sqrt{\frac{2}{\pi}}
\frac{\left(\ell+x\right)^2}{\lambda_D^3 s^{3/2}(1-s)^{3/2}}
\exp\left(-\frac{(\ell+ x)^2}{2 \lambda_D^2 s(1-s)}\right) \\
& \times &\!\!\!\! \Big\{ 1-\beta \left[V_1(\ell,s)+V_1(\ell,1-s)\right] -\beta \left[V_2(\ell,s)+V_2(\ell,1-s)\right]
\nonumber \\
&{}&+\frac{\beta^2}{2}
\left[V_1(\ell,s)+V_1(\ell,1-s)\right]^2  +o\left(\lambda_D^2\right) \Big\}.\nonumber
\end{eqnarray}
After computing the remaining integrals over $s$ and $\ell$  we eventually obtain 
\begin{eqnarray}\label{ro1_ht_bis}
\rho_{erg}(x,x,\beta) \!\!\!&=&\!\!\!\ \frac{e^{-\beta V(x)}}{\sqrt{2 \pi} \lambda_D}
\left[ 1 -\beta\frac{\mathrm{d^2} V}{\mathrm{d}x^2}
\frac{\lambda_D^2}{12}+\left(\beta\frac{\mathrm{d} V}{\mathrm{d}x}\right)^2 \frac{\lambda_D^2}{24}A_0 +o\left(\lambda_D^2\right)\right],
\end{eqnarray}
with
\begin{eqnarray}
A_0=-\frac{1}{2}&+&3\sqrt{2 \pi} \int_0^1 \mathrm{d} s\left(\frac{s}{1-s}\right)^{3/2}\int_0^{\infty}\mathrm{d}z\,z^2 \text{erfc}^2\left(\frac{z}{\sqrt{2 s}}\right)
\exp{\left( -\frac{z^2(2s-1)}{2s(1-s)}\right)}\nonumber \\&+&
3\sqrt{2 \pi} \int_0^1 \mathrm{d}s \int_0^{\infty}\mathrm{d}z\,z^2 \text{erfc}\left(\frac{z}{\sqrt{2 s}}\right)
\text{erfc}\left(\frac{z}{\sqrt{2(1-s)}}\right),
\end{eqnarray}
a numerical factor close 0.8 (estimated numerically to $0.796$).

\end{document}